# MoS$_2$ Quantum Dot/Graphene Hybrids for Advanced Interface Engineering of CH$_3$NH$_3$PbI$_3$ Perovskite Solar Cell with Efficiency over 20%

*Leyla Najafi,*[†,‡] *Babak Taheri,*[#,‡] *Beatriz Martín-García,*[†] *Sebastiano Bellani,*[†] *Diego Di Girolamo,*[#] *Antonio Agresti,*[#] *Reinier Oropesa-Nuñez,*[†,∥] *Sara Pescetelli,*[#] *Luigi Vesce,*[#] *Emanuele Calabrò,*[#] *Mirko Prato,*[⊥] *Antonio E. Del Rio Castillo,*[†] *Aldo Di Carlo,*[#,£*] *and Francesco Bonaccorso*[†,∥*]

[†] Graphene Labs, Istituto Italiano di Tecnologia, via Morego 30, 16163 Genova, Italy.

[#] C.H.O.S.E. (Centre for Hybrid and Organic Solar Energy), Department of Electronic Engineering, University of Rome Tor Vergata, Via del Politecnico 1, Rome 00133, Italy.

[∥] BeDimensional Srl., Via Albisola 121, 16163 Genova, Italy.

[⊥] Materials Characterization Facility, Istituto Italiano di Tecnologia, via Morego 30, 16163 Genova, Italy.

[£] L.A.S.E. - Laboratory for Advanced Solar Energy, National University of Science and Technology "MISiS", 119049 Leninskiy prosect 6. Moscow, Russia.

**ABSTRACT**

Interface engineering of organic-inorganic halide perovskite solar cells (PSCs) plays a pivotal role in achieving high power conversion efficiency (PCE). In fact, perovskite photoactive layer needs to work synergistically with the other functional components of the cell, such as charge transporting/active buffer layers and electrodes. In this context, graphene and related two-dimensional materials (GRMs) are promising candidates to tune "on demand" the interface properties of PSCs. In this work, we fully exploit the potential of GRMs by controlling the optoelectronic properties of hybrids between molybdenum disulfide ($MoS_2$) and reduced graphene oxide (RGO) as hole transport layer (HTL) and active buffer layer (ABL) in mesoscopic methylammonium lead iodide ($CH_3NH_3PbI_3$) perovskite ($MAPbI_3$)-based PSC. We show that zero-dimensional $MoS_2$ quantum dots ($MoS_2$ QDs), derived by liquid phase exfoliated $MoS_2$ flakes, provide both hole-extraction and electron-blocking properties. In fact, on the one hand, intrinsic n-type doping-induced intra-band gap states effectively extract the holes through an electron injection mechanism. On the other hand, quantum confinement effects increase the optical band gap of $MoS_2$ (from 1.4 eV for the flakes to > 3.2 for QDs), raising the minimum energy of its conduction band (from -4.3 eV for the flakes to -2.2 eV for QDs) above the one of conduction band of $MAPbI_3$ (between -3.7 and -4 eV) and hindering electron collection. The van der Waals hybridization of $MoS_2$ QDs with functionalized reduced graphene oxide (f-RGO), obtained by chemical silanization-induced linkage between RGO and (3-mercaptopropyl)trimethoxysilane, is effective to homogenize the deposition of HTLs or ABLs onto the perovskite film, since the two-dimensional (2D) nature of RGO effectively plug the pinholes of the $MoS_2$ QDs films. Our "*graphene interface engineering*" (GIE) strategy based on van der Waals $MoS_2$ QD/graphene hybrids enable $MAPbI_3$-based PSCs to achieve PCE up to 20.12% (average PCE of 18.8%). The possibility to combine quantum and chemical effects into GIE, coupled with the recent success of graphene and GRMs as interfacial layer, represents a promising approach for the development of next-generation PSCs.

# INTRODUCTION

Organic-inorganic halide perovskite solar cells (PSCs) are in the spotlight of the photovoltaic (PV) research to rival the leading technologies[1-4] (*i.e.* crystalline silicon solar cells[5-7] and inorganic thin-film solar cells[8,9]) since power conversion efficiency (PCE) exceeding 20%[10-12] can be obtained by affordable (low-cost and low-temperature) solution processing[13-15] with scaling-up prospective.[16-19]

Methylammonium lead iodide ($CH_3NH_3PbI_3$) perovskite ($MAPbI_3$) has been intensively studied as light harvesting material from the beginning of the PSC developments[1,2,20-22] and the elemental composition engineering of its chemistry[23,24] led to mixed cation and halide PSCs,[25] which boosted certified PCE above 22% (*i.e.* 22.1%[26] and 22.7%[10]). Although the archetypal $MAPbI_3$ reached certified maximum efficiency of 19.3%[27] (uncertified efficiency exceeding 20%[28-31]), it still covers a benchmarking role for the optimization and/or validation of PSC architectures[31-34] due to its historical breakthrough in the PV technology over the last years (starting from 2009, PCE of 3.8%[21]).[10] In particular, $MAPbI_3$-based PSCs provide a platform to study and design the interface between each functional layers of the PSCs,[16,35-41] whose carrier transport barrier determines undesirable hysteresis and instabilities effects in PSCs,[33,42-44] both in mesoscopic[31,45] and planar structure.[45] In fact, the photogenerated carriers have to be transported across the interfaces in the PSC structure,[38,46-48] and charge loss can occurs due to energy barriers and/or interfacial defects.[38,46-48] Therefore, appropriate energy level tailoring at the interfaces is pivotal in: 1) increasing open circuit voltage ($V_{oc}$); 2) facilitating charge transfer and extraction,[49,50] which contribute to increase short circuit current ($J_{sc}$) and fill factor (FF);[49,50] 3) eliminating hysteresis phenomena[51-53] and 4) extending lifetime[54,55] of current PSCs. In the run-up to reach the theoretical PCE limit (~31%)[56] of PSCs (practical limits of 29.5%[57] and 30.5%[58] have also been reported by considering intrinsic non-radiative recombination processes), graphene and related two-dimensional (2D) materials (GRMs) are emerging as a paradigm shift of interface engineering to boost the PV performance.[56,59-73] Actually, the large variety of GRMs offers peculiar (opto)electronic properties[74,75] that can be on-demand tuned by means of morphological modification[76,77] and chemical functionalization.[78-80] Moreover, GRMs can be produced from the exfoliation of their bulk counterpart in suitable solvents[81-85] in form of functional inks,[86] which can be deposited on different substrates by established large-scale, cost-effective printing/coating techniques,[87-90] compatible with solution-based manufacturing of PSCs.[15,91-93]

With the aim to deeply exploit the use of 2D materials for engineering the interface of PSCs, herein we report a synergistic quantum-chemical approach for controlling the energy band levels and the thin-film morphology of low dimensional, van der Waals hybrids between molybdenum disulfide QDs ($MoS_2$ QDs) and reduced graphene oxide (RGO) as hole transport layer (HTL) or active buffer layer (ABL) (between HTL and Au electrode) in mesoscopic $MAPbI_3$-based PSCs (**Figure 1**a). Notably, both 2D $MoS_2$ and RGO have been previously reported as possible HTLs[64,72,94-103] or ABL.[64,104,105] However, their intrinsic work function (WF) (typically < 4.8 eV for both pristine $MoS_2$[106-109] and RGO[110-113]) is inferior to that of conventional HTL materials, including 2,2',7,7'-Tetrakis-(N,N-di-4-methoxyphenylamino)-9,9'-spirobifluorene (spiro-OMeTAD) (WF > 4.9,[114,115] especially for the doped forms mostly exploited as HTLs[114-117]) and poly(3,4-ethylenedioxythiophene):poly(styrene sulfonate) (PEDOT:PSS) (WF ranging between 5.0 and 5.2 eV[118-121]). This can limit the hole extraction process.[95,122,123] Moreover, the optical band gap ($E_g$) of both RGO (< 2 eV,[124,125] depending on its oxidation level[124,125]) and $MoS_2$ (~1.2 eV for bulk,[126,127] and ~1.8 eV for single-layer[128-130]) results in minimum energy of conduction band (CB) (~-4.3 eV[131-133]) lower than that of lowest unoccupied molecular orbital (LUMO) (reported between -4.0[134-137] and -3.7 eV[114,138,139]), not providing electron blocking properties.[140,141] Therefore, physical/chemical modification of 2D $MoS_2$ and RGO are needed to tune the optoelectronic properties for their efficient implementation as HTL.[95,99,110,122,123,142,143] As shown in Figure 1b, zero-dimensional (0D) anodic interlayer of $MoS_2$ quantum dots ($MoS_2$ QDs), derived by liquid phase exfoliation (LPE) of $MoS_2$ flakes,[144] hold optimal electronic structure to effectively extract the photogenerated holes through electron injection mechanism[145-147] from their intrinsic n-type doping[148-151]-induced intra-band gap states.[152,153] The latter have been reported to be a consequence of the natural presence in $MoS_2$ of sulfur vacancies,[153-

[158] impurities[159,160] and defect.[161-164] Quantum confinement effects open the MoS$_2$ optical bandgap (from 1.4 eV for the flakes to > 3.2 eV for the QDs), raising the minimum energy of the CB of MoS$_2$ (from -4.3 eV for the flakes to -2.2 eV for the QDs) above the energy of LUMO of MAPbI$_3$ (between -4.0[134-137] and -3.7 eV[114,138,139]), thus providing electron-blocking properties. Hole-extraction and electron-blocking properties of MoS$_2$ QDs synergistically suppress the interfacial recombination losses observed in benchmark devices (fluorine doped tin oxide (FTO)/compact TiO$_2$ (cTiO$_2$)/mesoporous TiO$_2$ (mTiO$_2$)/MAPbI$_3$/2,2',7,7'-Tetrakis-(N,N-di-4-methoxyphenylamino)-9,9'-spirobifluorene (spiro-OMeTAD)/Au),[34,165] and in previous cell architectures exploiting native MoS$_2$ flakes as ABLs.[64,165] With the aim to form homogeneous (*i.e.* pinhole-free) nm-thick HTLs, MoS$_2$ QDs, which do not cover totally the MAPbI$_3$ film after their deposition, are hybridized with chemically (3-mercaptopropyl)trimethoxysilane (MPTS)-functionalized RGO (f-RGO) flakes[166] (the resulting hybrid is herein named MoS$_2$ QDs:f-RGO).

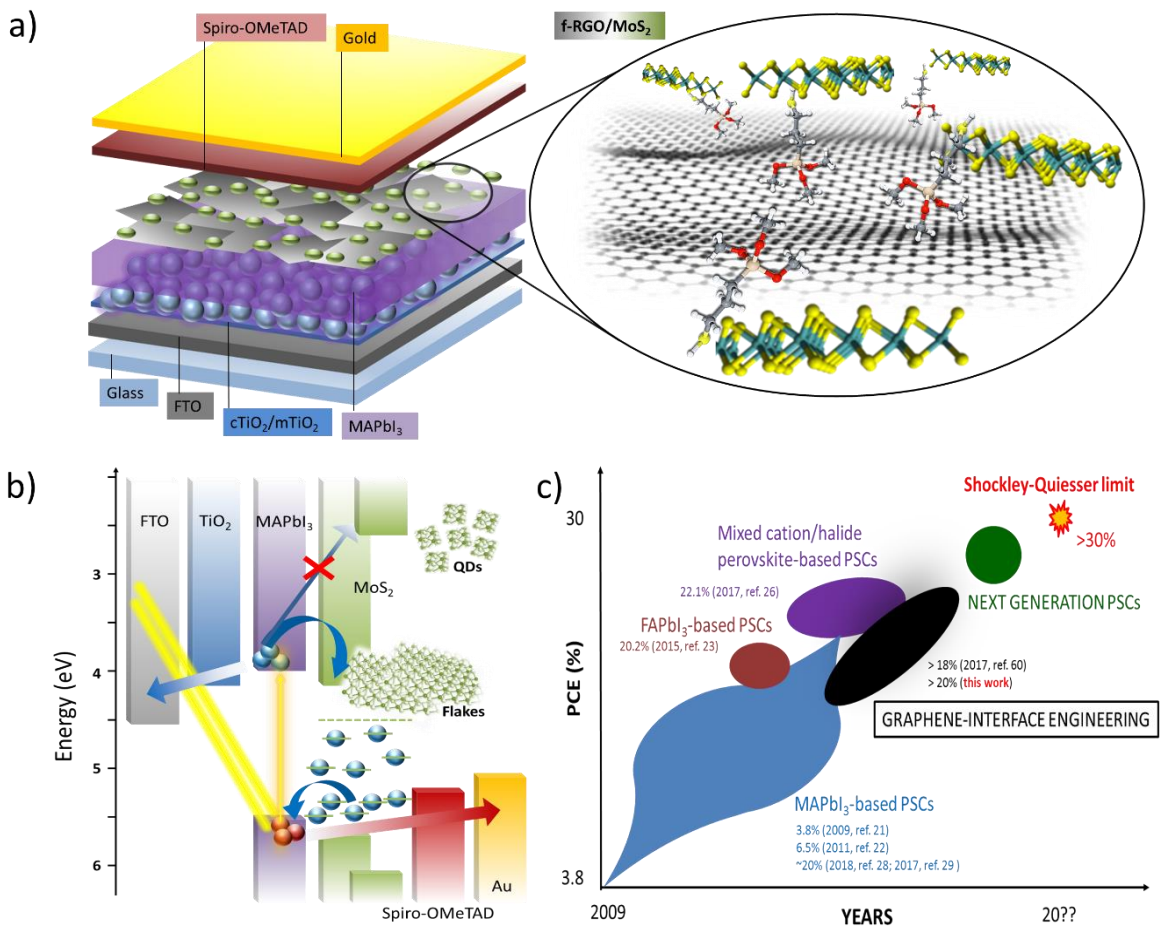

**Figure 1.** (a) Sketch of mesoscopic MAPbI$_3$-based PSC exploiting MoS$_2$ QDs:f-RGO hybrids as both HTL and ABL. (b) Scheme of the energy band edge positions of the materials used in the different components of the assembled mesoscopic MAPbI$_3$-based PSC. The energy band edge positions of MoS$_2$ flakes and MoS$_2$ QDs were determined from OAS and UPS measurements detailed along the text, while those of the other materials were taken from literature: FTO,[52] TiO$_2$,[52] MAPbI$_3$,[134-139] spiro-OMeTAD[52] and Au[52]. (c) State-of-the-art and predicted PCE evolution for PSCs, highlighting the synergistic potential of GIE and the formulation of advanced perovskite chemistries.

The RGO flakes are effective to plug the pinholes MoS$_2$ QDs films, thus to homogenize the HTL. The choice of the functionalization for RGO relies on the bifunctional role of MPTS molecules, which effectively anchor onto the RGO flakes (*via* silanization-mediated bonding),[167,168] while exposed thiol (SH) moieties enable f-RGO to interact with MoS$_2$ QDs (*via* S-S van der Waals physisorption[169] and/or S-vacancies passivation/filling[170,171]). Our results show the potential of quantum and chemical effects into "*graphene interface engineering*" (GIE) to produce highly performant MAPbI$_3$-based PSCs with PCE up to 20.12% (average PCE of 18.8%). The remarkable advances achieved also in the exploitation of graphene flakes- and graphene QDs-doped ETLs, including both mesoscopic TiO$_2$[45] and solution-processed SnO$_2$[28], makes GIE a versatile tool for the design of record-high efficiency (solution-processed) next-generation PSCs (Figure 1c).

**RESULTS AND DISCUSSION**

**Production and characterization of MoS$_2$ QDs, f-RGO and MoS$_2$ QDs:f-RGO**

The MoS$_2$ QDs were produced through a facile and scalable one-step solvothermal approach starting from MoS$_2$ flakes previously obtained by LPE of bulk MoS$_2$ in 2-propanol (IPA),[144] followed by sedimentation-based separation (SBS) process.[172,173] The LPE process exploits hydrodynamic shear-forces-controlled ultrasonication to overcome the van der Waals forces (15-20 meV Å$^{-2}$,[174] or ~5 meV/atom,[175,176]) that bind MoS$_2$ layers.[81,84,177,178] The SBS process separates various particles on the basis of their sedimentation rate in response to a centrifugal force acting on them.[84,179,] Consequently, MoS$_2$ flakes were first produced by LPE and subsequently solvothermally treated for the production of MoS$_2$ QDs.[144] Then, by exploiting SBS we selected MoS$_2$ QDs, while residuals MoS$_2$ flakes were discarded as sediment.[144] Reduced graphene oxide was produced by thermal annealing (1000 °C under a 100 sccm flow of Ar (90%):H$_2$ (10%)) of graphene oxide (GO)[180,181] synthetized from graphite flakes using a modified Hummer's method.[182] Subsequently, RGO was functionalized by MPTS *via* silanization-mediated chemical bonding.[166,183] The silanization process was triggered by the hydrolization and condensation of the methoxy groups (–OCH$_3$) of MPTS, which react with the O moieties of RGO (**Scheme 1**a).[166,167] A solvent-exchange process[95,184,185] was carried out to re-disperse MoS$_2$ QDs and f-RGO dispersions in IPA. The hybrid MoS$_2$ QDs:f-RGO dispersion was obtaining by mixing f-RGO and MoS$_2$ QDs dispersions with a material weight ratio of 1:2. The hybridization of the materials is completed by the exposed SH moieties of f-RGO, which interact with MoS$_2$ QDs *via* S-S van der Waals physisorption[169] and/or passivation/filling of the S-vacancies of MoS$_2$ QDs[170,171] (Scheme 1b).

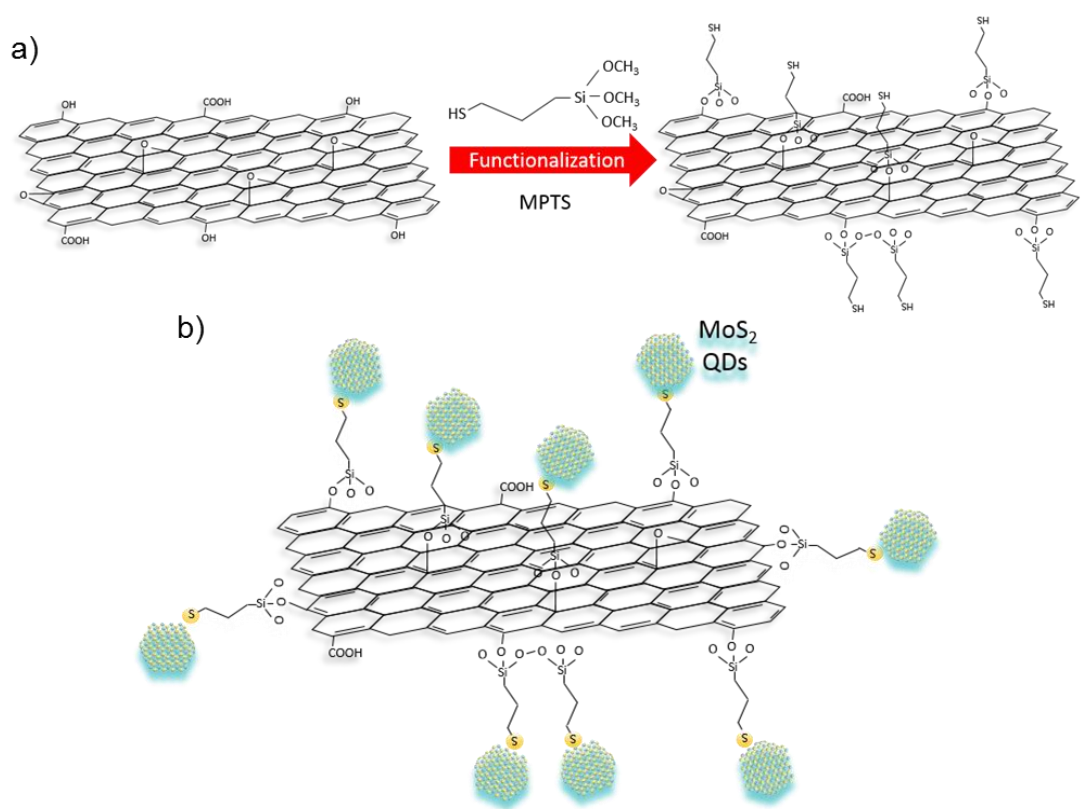

**Scheme 1.** Representative sketches of: (a) the functionalization of RGO (structure based on the Lerf-Klinowski model)[186] with MPTS; (b) the hybridization of MoS$_2$ QDs and f-RGO.

The details of material production and dispersion formulation are reported in Experimental Method section. It is worth noting that the final material dispersions were obtained in low-boiling point alcohol-based solvents, which are compatible with environmentally friendly, low-temperature and solution-processed deposition methods. By taking advantage of this approach, we exploited spray coating deposition, since it can be applied on irregular surface with higher reproducibility than that obtained with other deposition methods (*e.g.,* spin/blade coating and screen printing).[187] Moreover, with a broader context vision, spray coating is a promising technique to speed up the production of perovskite modules fabrication[188,189] in view of their market entry.[190,191]

The lateral size and thickness of the as-produced $MoS_2$ QDs and f-RGO samples were evaluated by means of transmission electron microscopy (TEM) and atomic force microscopy (AFM), respectively. **Figure 2**a,b show representative TEM and AFM images of $MoS_2$ QDs. Microscopy statistical analysis of lateral dimension (Figure 2c) and thickness (Figure 2d) shows lognormal distributions peaking at ~2.6 nm and ~1.6 nm, respectively, which means that both one- and few-layer QDs were effectively produced (the monolayer thickness is between 0.7 and 0.8 nm[128,192]). Notably, the thickness distribution of $MoS_2$ QDs is similar to that measured for the native $MoS_2$ flakes (average thickness of ~2.7 nm), whose morphological characterization is reported in the Supporting Information (**Figure S1**). Figure 2e,f report TEM and AFM images of f-RGO flakes, which exhibited irregular shape and rippled paper-like morphology. Microscopy statistical analysis of lateral dimension (Figure 2g) and thickness (Figure 2h) displays lognormal distribution peaked at ~980 nm and ~1.3 nm, respectively. These values are comparable with those obtained for native RGO (average lateral dimension and thickness of 1.7 μm and 1.8 nm, respectively)[183] (see **Figure S2**). Since the thickness of single-layer pristine graphene is ~0.34 nm,[193,194] these data indicated that our methodology produced few-layer f-RGO flakes. The structural properties of the materials were investigated by Raman spectroscopy (see SI for Raman spectroscopy analysis additional details, **Figure S3**), confirming their exfoliated crystal structure.

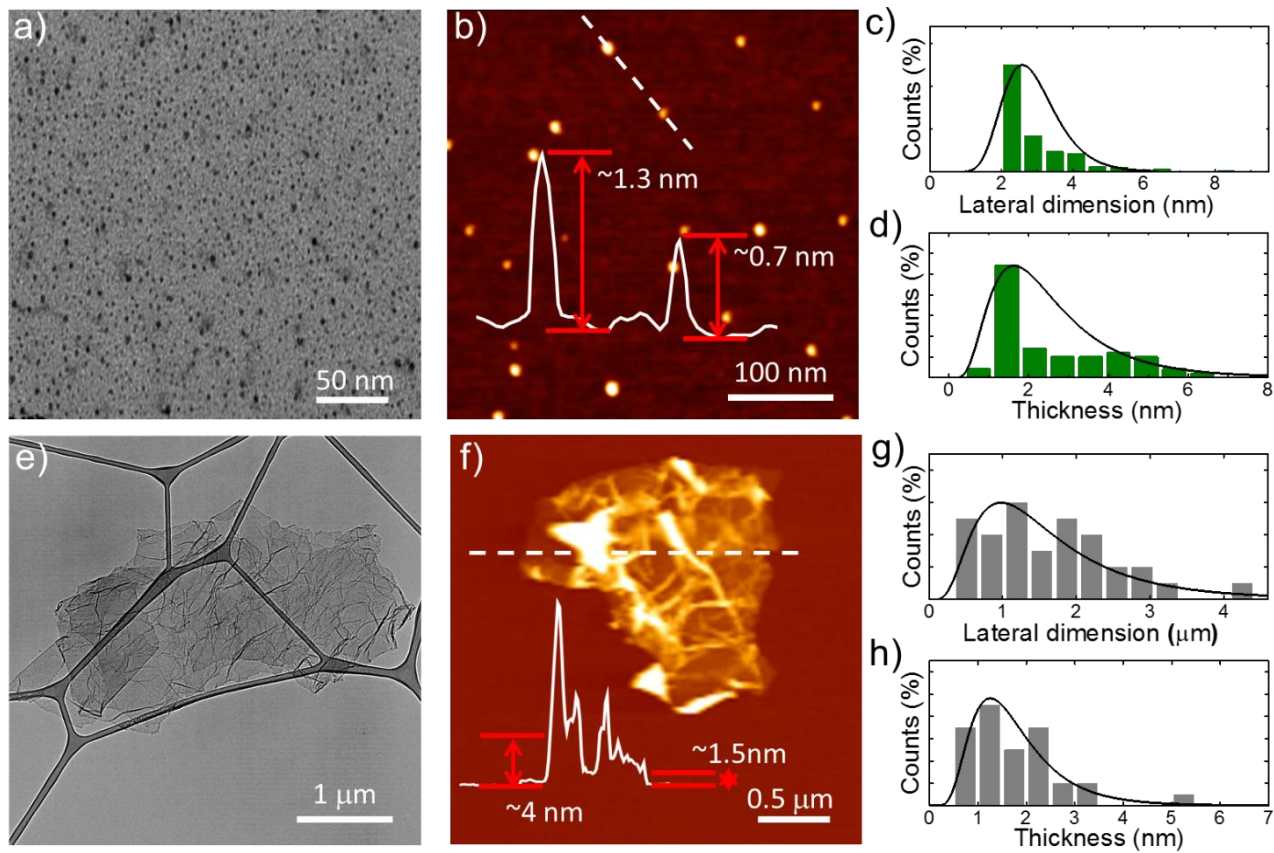

**Figure 2.** Morphological characterization of (a-d) the as-produced MoS$_2$ QDs and (e-h) f-RGO flakes. TEM and AFM images of (a, b) MoS$_2$ QDs and (e, f) f-RGO and the corresponding statistical analysis of (c, g) the lateral dimension and (d, h) the thickness, respectively. The statistical analyses were performed on 50 flakes from the different images collected. Representative height profiles (solid white lines) of the indicated sections (white dashed lines) are also shown in panels (b) and (f).

X-ray photoelectron spectroscopy (XPS) measurements were carried out to determine the elemental composition, chemical phase and interaction of the as-produced MoS$_2$ QDs:f-RGO samples. **Figure 3**a shows the S 2s and Mo 3d XPS spectra of MoS$_2$ QDs:f-RGO, together with their deconvolution. The peaks at the lowest binding energy (~226 eV and ~228 eV) and the peak at ~229 eV are assigned to S 2s[144,195,196] and Mo 3d$_{5/2}$, respectively, of the semiconducting 2H (trigonal prismatic) phase of MoS$_2$.[144,195,196] The peak centered at ~232.5 eV can be fitted with two components.[64,144] The first component (~232 eV) is assigned to Mo 3d$_{3/2}$ of the 2H-MoS$_2$. Instead, the second one (~233 eV), as well as the low intensity peak centered at ~236 eV, are associated with oxidized phases of Mo,[177,197] which can be produced during the LPE of bulk MoS$_2$.[144,177] However, the percentage content (%c) of oxidized Mo (< 3%) indicates that only a small fraction of the material is oxidized.

Noteworthy, the initial LPE step in IPA can overcome the drawbacks of the LPE of transition metal dichalcogenides[144,198] (including $MoS_2$[144]) in conventional high-boiling point solvents, such as N-methyl-2-pyrrolidone (NMP) [64,177,178,199] and N,N-dimethylformamide (DMF), [64,177,178] where high percentage content (%c) of oxidized species (between 40% and 60%, depending on the process parameter) are typically produced.[64,177] Moreover, the functionalization process did not cause any additional oxidation of $MoS_2$ QDs, since the %c of oxidized Mo in the hybrids (< 3%) is also inferior to that of both $MoS_2$ flakes (< 7%)[144] and $MoS_2$ QDs (~ 5%)[144]. The component at ~228 eV appeared in the S 2s XPS spectrum is ascribed to the presence of SH moieties of f-RGO and proves the interaction of the SH moieties with $MoS_2$ QDs *via* S-S van der Waals physisorption[169] and/or passivation/filling of the S-vacancies of $MoS_2$ QDs,[170,171] as proposed in Scheme 1b. The ratio between %c of Mo 3d and S 2s ascribed to $MoS_2$ QDs is ~0.5, while for the $MoS_2$ QDs the same ratio was > 0.5.[144] This confirms that the functionalization process restored the S-vacancies of the native $MoS_2$ flakes, in agreement with the hybridization mechanism reported in Scheme 1b.

Figure 3b shows the C 1s spectrum of $MoS_2$ QDs:f-RGO, which is deconvoluted in six components, indicating the C $sp^2$ and $sp^3$ network and oxygen functionalities of the f-RGO.[200] The C $sp^2$ component, centered at 284.5 eV, dominates the spectrum (%c ~75.7%), indicating that the delocalized π-conjugated structure was almost fully restored after the thermal reduction of the native GO (%c of C $sp^2$ ~48.5%[183]).[201,202] Noteworthy, the %c of C $sp^2$ in f-RGO is almost the same of that of native RGO (75.6%[183]), indicating that the functionalization process do not alter the basal properties of the RGO flakes.[183] The component peaking at ~286.9 eV is assigned to C-O bonds (%c = 6.9%). A residue of C $sp^3$ is still present (peak centered at 284.7 eV, %c ~7.6%) as well as carboxylate carbon O–C=O bonds (peak centered at 287.7 eV, %c ~1%). The component observed at ~290.7. eV (%c ~3.7%) is attributed to π-π* satellite structure (extended delocalized electrons),[203,204] a characteristic of aromatic C structure. The component peaking at 284.0 eV is also significantly present (%c ~14%). This band can arise as a consequence of C lattice vacancies/distortions introduced mainly during the GO reduction process[183] and/or the hybridization of f-RGO with MoS$_2$ QDs. Regarding the MPTS functionalization of the RGO, Figure 3c shows the Si 2s and S 2p XPS spectra, together with their deconvolution. The appearance of the silane and SH-doublet components peaking at ~153.5 eV and ~163.4 eV, respectively, [166,205] are fingerprints of the MPTS. The components peaking at ~168.4 eV and ~169.7 eV are attributed to S 2p doublet of SO$_4^{2-}$.[166,206] These oxidized species are due to both MPTS interacting with O moieties on RGO flakes[167,207] and MPTS oxidized during the functionalization process. The %c of the oxidized groups is ~5% of the total S content. The ratio between the sum of the %c of SH free and S-S bnds related to the MPTS and that of C bonds on f-RGO is 3%, and estimate the percentage extent of the functionalization of RGO with MPTS.[183] The interaction between the MPTS and the RGO flakes was confirmed by complementary Fourier-transform infrared (FTIR) measurements. After RGO functionalization, in the f-RGO FTIR spectrum (see SI, **Figure S4**), the Si-O-Si stretching band appears at 1078 cm$^{-1}$ shifted and broaden compared to pure MPTS (1089 cm$^{-1}$), indicating the coupling between the alkoxy silane groups and the oxygen groups of RGO, as proposed in Scheme 1. Even more, Si-O-Si band was still present in MoS$_2$ QDs:f-RGO keeping the same position. This indicated that no modification occurs in the silane-RGO interaction path, leaving only the SH groups as linker option for the MoS$_2$ QDs (see SI for further discussion). The effectiveness of the MPTS functionalization was also macroscopically observed by noting the improved dispersibility of f-RGO in ethanol compared to that of RGO (**Figure S5**). In fact, MPTS are polar molecules able to decrease the surface energy of native RGO in alcohol-based solvents (~46.1 mN m$^{-1}$ in ethanol),[208,209] enhancing dispersion stability and hindering formation of aggregates during films deposition.[183] In addition, XPS analysis also evidences that S 2p doublet related to S-S bonds (centered at ~164.5 eV) create an interconnection between the MPTS molecules and MoS$_2$ QDs, since the corresponding %c increases from 10% of the total S content in f-RGO[167,205] to ~21% of the total S content in MoS$_2$ QDs-f-RGO.

Optical absorption spectroscopy (OAS) and ultraviolet photoelectron spectroscopy (UPS) measurements were carried out in order to assess the charge-extraction/blocking capability of MoS$_2$ QDs. Native MoS$_2$ flakes were also measured for comparison, since they have been previously reported as effective ABL between MAPbI$_3$ and spiro-OMeTAD in mesoscopic PSCs.[64] The UV-Vis absorption spectrum of MoS$_2$ flakes (Figure 3d) shows peaks at ~670 nm and ~620 nm, which arise from the excitonic transitions between the split valance bands and the minima of the conduction band at the *K*-point of the Brillouin zone of layered MoS$_2$,[210,211] known as the A and B, respectively. More in detail, the spin-orbit interaction and interlayer coupling are responsible for the valence band (VB) splitting. The energy difference between the A and B is ~180 meV, which agrees with the values predicted by density functional theory (DFT) calculations (146 meV and 174 meV for monolayer and bilayer, respectively).[212] The broad absorption band centered at ~400 nm arises from the C and D inter-band transitions between the density of state peaks in the valence and conduction bands.[213,214] Differently, MoS$_2$ QDs do not show the absorption peaks of MoS$_2$ flakes, and their absorption edge shifts toward lower wavelength compared to the latter. This is a consequence of quantum 0D-confinement,[215,216] which affects the optical properties of nanostructures when their size is comparable or smaller than the excitonic Bohr radius (~23 nm for MoS$_2$).[217] Quantum confinement, as well as edge effects, endow excitation-dependent photoluminescence (PL) properties in MoS$_2$ QDs, (**Figure S6**).[144,215,218,219] In fact, the PL emission peak of MoS$_2$ QDs was red-shifted with increasing excitation wavelength.[144] In order to further confirm the effect of quantum confinement on the optical properties of MoS$_2$, the E$_g$ was evaluated by the (αhν)$^n$ *vs.* hν (Tauc plot) analysis (Figure 3e) using the Tauc relation Ahν = Y(hν-E$_g$)$^n$, where A is the absorbance, *h* is the Planck's constant, ν is the photon's frequency, and Y is a proportionality constant.[220] The value of the exponent denotes the nature of the electronic transition, discriminating between direct-allowed transition (*n* = 2) or indirect-allowed transition (*n* = 0.5).[221] Bulk MoS$_2$ is an indirect bandgap semiconductor with E$_g$ = 1.29 eV.[222] With decreasing thickness, theoretical[223,224] and experimental[128] studies revealed a progressive confinement-induced shift in the indirect bandgap from the bulk value of 1.29 eV up to 1.90 eV, while the direct bandgap increases by only 0.1 eV.[128] As a consequence of these different scaling properties, MoS$_2$ undergoes a crossover from an indirect bandgap semiconductor to a direct bandgap material in the monolayer limit.[128,223,224] The indirect-to-direct bandgap transition has a strong impact on the PL emission, which shows a dramatic enhancement (by more than a factor of 1000) compared to the one of the bulk counterpart.[225] In the case of our $MoS_2$ flakes, PL was not detected, in agreement with a dominant few-layer nature evidenced by the AFM statistical analysis of the thickness (Figure 2h). Consequently, in Tauc analysis *n* was set equal to 0.5 (indirect-allowed transition). Differently, $MoS_2$ QDs, although show similar thickness of the $MoS_2$ flakes (see Figure 2d,h), display remarkably PL, since additional quantum effects arising from 0D-confinement activate direct bandgap behavior,[215,216,217] as proved by PL emission (Figure S6). Then, in Tauc analysis *n* was set equal to 2 (direct-allowed transition). It is worth noting that *n* undergoes a size (*i.e.* thickness and lateral dimension)-dependent transition from 2 in direct-bandgap bulk semiconductors to 1 in direct-bandgap nanocrystal.[226] Consequently, the calculated $E_g$ value of $MoS_2$ QDs has to be considered qualitatively. Taking into account these consideration, the estimated $E_g$ increase from ~1.4 eV for $MoS_2$ flakes to ~4.0 eV (~3.2 eV assuming *n* = 1, **Figure S7**) for $MoS_2$ QDs, in agreement with previous studies.[227]

Ultraviolet photoelectron spectroscopy measurements allowed the energy Fermi level ($E_F$), *i.e.* the WF, and VB to be determined. Figure 3f shows that secondary electron cut-off (threshold) energies of the He-I (21.22 eV) UPS spectra is the same for $MoS_2$ flakes and QDs (~16.8 eV), corresponding to a WF of ~4.6 eV. The inset to Figure 3f shows the UPS spectra region near the $E_F$, which allows the maximum energy of VB to be estimated at ~-5.7 eV for $MoS_2$ flakes and ~-6.2 eV for $MoS_2$ QDs. Taking into account the $E_g$ values estimated by Tauc analysis, the minimum energy of conduction band (CB) is estimated at ~-4.3 for $MoS_2$ flakes and ~-2.2 eV for $MoS_2$ QDs. The results indicate that, differently from $MoS_2$ flakes, $MoS_2$ QDs have minimum energy of CB lower than that of LUMO of $MAPbI_3$ (between -4.0[134-137] and -3.7 eV[114,138,139]). Consequently, $MoS_2$ QDs effectively act as electron blocking material into the PSC structures (see Figure 1b). Furthermore, the UPS data revealed that both $MoS_2$ flakes and QDs cannot collect hole from their VB, since the corresponding energy levels (~-5.7 eV for $MoS_2$ flakes and ~-6.2 eV for $MoS_2$ QDs) are inferior to the one of $MAPbI_3$ (~-5.4 eV[137-139]). Since $MoS_2$ flakes have been previously reported as HTL material, [64,72,94-

[99] the holes can be extracted from the MAPbI$_3$ by injecting electrons[145-147] from inter-gap states of MoS$_2$.[152,153] The latter have been reported to be a consequence of the intrinsic presence in MoS$_2$ of S-vacancies[153-158], impurities[159,160] and defects.[161-164] The presence of these inter-gap states is confirmed by UPS data, which reveal an intrinsic n-type doping of MoS$_2$ flakes ($E_F$ is just 0.3 eV inferior to the energy of energy minimum of CB, and 1.1 eV superior to the energy maximum of VB). This observation is also in agreement with previous studies,[148-151] which evidenced n-type-doping transport measurements in MoS$_2$-based field effect transistor.[148-151] Since XPS analysis revealed equal stoichiometry between MoS$_2$ flakes and MoS$_2$ QDs, the hole extraction mechanism deduced for MoS$_2$ flakes can also be valid for MoS$_2$ QDs.[144] We also point out that a similar hole-extraction mechanism has been reported for MoO$_3$ anodic interlayer,[146,147,228] due to inherent n-type behavior that allows the material to act as donors in transparent conducting oxides.[229]

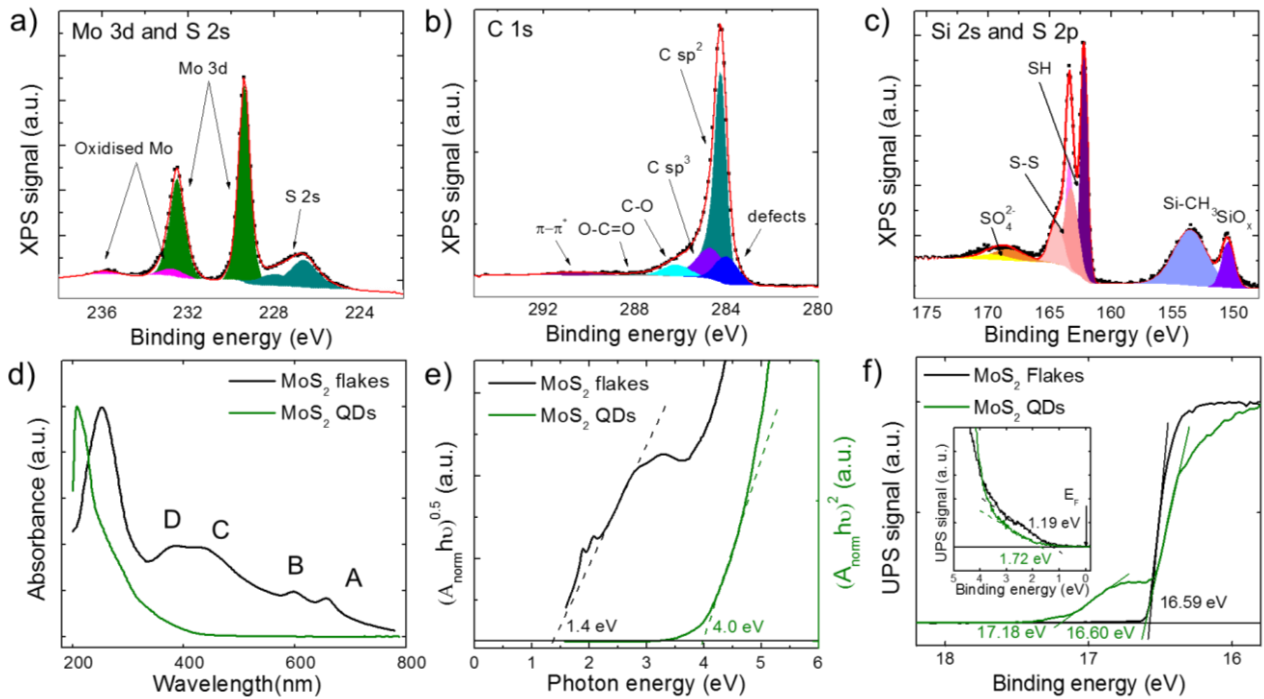

**Figure 3.** (a) Mo 3d and S 2s, (b) C 1s and (c) Si 2s and S 2p XPS spectra for MoS$_2$ QDs:f-RGO. The deconvolution of the corresponding XPS spectra is also shown. (d) Absorption spectra of MoS$_2$ flakes and QDs. (e) Tauc plots of MoS$_2$ flakes and QDs. (f) Secondary electron threshold region of He-I UPS spectra of MoS$_2$ flakes and QDs, which were used for estimating the WF values. The inset shows VB region of He-I UPS spectra of MoS$_2$ flakes and QDs which were used for estimating $E_F$ values.

**Perovskite solar cell architecture and their characterization**

The optoelectronic characterization of MoS$_2$ QDs evidenced that they hold both hole-extraction and electron blocking properties, which are the most important requirements for application as HTL.[41,40,230] In order to deposit continuous and homogeneous film based on MoS$_2$ QDs, the latter were hybridized with f-RGO flakes, whose 2D nature spontaneously plugs the pinholes in MoS$_2$ QD films. With the aim to prove the effectiveness of MoS$_2$ QDs or MoS$_2$ QDs:f-RGO as hole transport materials, both of them were incorporated into mesoscopic MAPbI$_3$-based PSCs to be used as HTL or ABL between MAPbI$_3$ and spiro-OMeTAD. The investigated PSC has the following architecture FTO/cTiO$_2$/mTiO$_2$/MAPbI$_3$/ABL (MoS$_2$ QDs or f-RGO or MoS$_2$ QDs:f-RGO)/spiro-OMeTAD/Au. The ABL were also tested as HTL in absence of spiro-OMeTAD. Additional details of the device fabrication are reported in Experimental Methods. It is important to note that MoS$_2$ QDs and MoS$_2$ QDs:f-RGO nm-thick films were deposited onto MAPbI$_3$ films by spray coating the respective dispersions in IPA. The compatibility of the IPA with the MAPbI$_3$ layer was assessed in previous experiments.[64,231] A representative FTO/cTiO$_2$/mTiO$_2$/MAPbI$_3$/MoS$_2$ QDs:f-RGO/spiro-OMeTAD/Au architecture was characterized by cross-sectional scanning electron microscopy (SEM) (**Figure 4**a). The MoS$_2$ QDs:f-RGO layer is not resolved because of its nm-scale thickness. However, top-view SEM analysis (Figure 4b,c) of MAPbI$_3$ surface shows that the contrast between different MAPbI$_3$ grains is reduced after the deposition of ultrathin layer of MoS$_2$ QDs:f-RGO. This indicates the presence of a nm-thick film of MoS$_2$ QDs:f-RGO covering MAPbI$_3$. In addition, SEM/elemental analysis by energy-dispersive X-ray spectroscopy (EDX) were performed on the different MAPbI$_3$-based PSCs (**Figure S8**), focusing on Pb (M, 2.34 keV), Mo (L$\alpha$, 2.29 keV) and C (K$\alpha$, 0.28 keV) peak signal, to evaluate the coverage of the MAPbI$_3$ surface with MoS$_2$ QDs and f-RGO. After the deposition of the MoS$_2$ QDs:f-RGO, a significant increase of the C signal relative to Pb was observed (C/(Pb+Mo) atomic ratio = 15±2) compared to the reference device (without spiro-OMeTAD, C/Pb atomic ratio = 2.2±0.1). This further confirms the coverage of the MAPbI$_3$ surface by the MoS$_2$ QDs:f-RGO film. Similar results were also evidenced by using only f-RGO (C/Pb atomic ratio = 16±2).

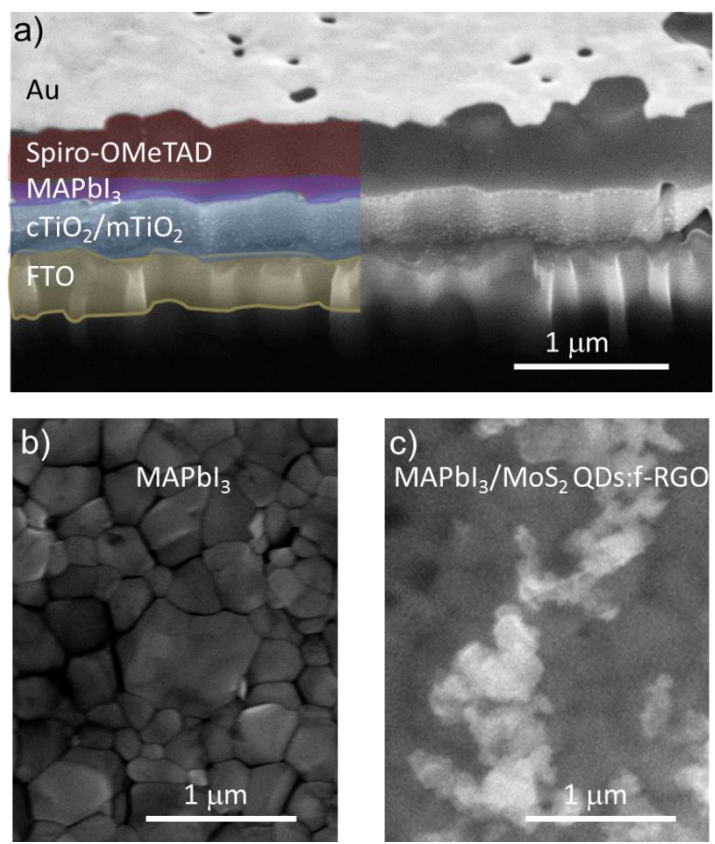

**Figure 4.** Morphological SEM characterization of PSCs. (a) Cross-sectional SEM image of a representative FTO/cTiO$_2$/mTiO$_2$/MAPbI$_3$/MoS$_2$ QDs:f-RGO/spiro-OMeTAD/Au architecture. Top-view SEM images of (b) MaPbI$_3$ surface and (c) MoS$_2$ QDs:f-RGO film deposited onto MAPbI$_3$ layer.

The capability of MoS$_2$ QDs-, f-RGO- and MoS$_2$ QDs:f-RGO-based ABLs to effectively collect the photogenerated holes was proved by measuring the PV performance of PSCs without spiro-OMeTAD. The current-voltage (I-V) characteristics (**Figure 5**a) show that the best performances are obtained for PSCs adopting MoS$_2$ QDs:f-RGO as HTL. Such PSCs exhibited a significant increase of PCE compared to the ones without HTL (7.60% *vs.* 3.01%). Also MoS$_2$ QDs and f-RGO individually enhanced the PV performance of the HTL-free reference. In particular, MoS$_2$ QDs increased the J$_{sc}$ of the HTL-free reference from 10.33 mA cm$^{-2}$ to 15.52 mA cm$^{-2}$, while f-RGO boosted the V$_{oc}$ of the HTL-free reference from 0.69 V to 0.81 V. The poor V$_{oc}$ obtained by the MoS$_2$ QDs is ascribed to a current leakage in absence of complete coverage of the MAPbI$_3$ surface, leading to charge recombination at the MAPbI$_3$/Au interface, since the metallic behavior of Au is not hole-selective.[232,233] This drawback can be overcome by hybridizing MoS$_2$ QDs with f-RGO, whose planar nature can provide an effective coverage of the MAPbI$_3$. These data indicate that the individual components of MoS$_2$ QDs:f-RGO synergistically improve the PV performance of the cells. Although the measured PV performance are still lower compared to PSC based on conventional HTLs, these results are promising for the development of viable alternative HTLs based on GRMs.

After these preliminary tests, MoS$_2$ QDs, f-RGO and MoS$_2$ QDs:f-RGO were tested as ABL between MAPbI$_3$ and spiro-OMeTAD. As shown by the I-V curves of representative PSCs in Figure 5b, the PV performance increased with the addition of f-RGO and MoS$_2$ QDs:f-RGO as ABLs compared to the reference device. The "champion cell" using MoS$_2$ QDs:f-RGO reached a maximum PCE of 20.12%., a V$_{oc}$ of 1.11 V, a J$_{sc}$ of 22.81 mA cm$^{-2}$, and a FF of 79.75%. The reference device has shown a PCE of 16.85%, with a V$_{oc}$ 1.07 V, a J$_{sc}$ of 20.28 mA cm$^{-2}$, and a FF of 76.9%. The device using only the MoS$_2$ QDs as ABL reached PCE of only 14.40%, thus without improving the PV performance of the reference one. The incorporation of f-RGO remarkably increases the J$_{sc}$ of the reference device up to 22.76 mA cm$^{-2}$, reaching a PCE of 18.64%. The enhanced J$_{sc}$ of the PSCs exploiting f-RGO and MoS$_2$ QDs:f-RGO compared to the value obtained by both the reference device and the one based on MoS$_2$ QDs as ABL is attributed to the efficient charge collection in presence of f-RGO and MoS$_2$ QDs:f-RGO, respectively. Notably, the optical absorption of the MoS$_2$ QDs:f-RGO-based PSCs does not show significant differences compared to that of MoS$_2$ QDs-based PSC, showing an increase of only ~3% and ~8% compared to f-RGO-based and reference PSCs, respectively (**Figure S9**). Hysteresis phenomena, such as anomalous dependence on the voltage scan direction/rate/range,[43,234] voltage conditioning history,[235] and device configuration,[36] could affect the I-V measurements.[236,237] In order to exclude such effects, the PCE over time at the maximum power point (MPP) was measured for a different batch of cells (**Figure S10**), confirming that the MoS$_2$ QDs:f-RGO improves the PSC performance of the reference PSC. Forward and reverse I-V curves were also collected (see comparative results of the different ABLs tested SI, **Figure S11)**, showing that the presence of MoS$_2$ QDs:f-RGO as ABLs decreases the hysteresis phenomena compared to those of the other investigated PSCs, including the reference device adopting spiro-OMeTAD as HTL without ABL.

Steady-state PL measurements were performed to evaluate the capability of the ABLs to extract the photogenerated holes from the MAPbI$_3$. In fact, the hole-extraction process hinders the radiative charge recombination in the absorber material,[238-240] which then show a PL quenching.[241,242] Figure 5c shows that the addition of ABL between MAPbI$_3$ and spiro-OMeTAD suppressed the PL emission of MAPbI$_3$. Quantitatively, the PL decreased by 49.5%, 51.9% and 65.8% in presence of f-RGO, MoS$_2$ QDs, and MoS$_2$ QDs:f-RGO, respectively. This means that the ABLs accelerated the hole-extraction dynamics at the photoelectrode.[239,243] However, I-V measurements clearly show the need of f-RGO to increase the $J_{sc}$ indicating that other effects, such as the morphology of the ABL films, practically influence the PV performance of the PSCs. Incident power conversion efficiency (IPCE) measurements (Figure 5d) are consistent with the I-V ones. In fact, they show that f-RGO and MoS$_2$ QDs-f-RGO increased the IPCE in the 350-750 nm range by ~5% and ~7% compared to the one of the reference device and the MoS$_2$ QDs-based PSCs, respectively. The trend of integrated current density ($J_{IPCE}$) values calculated from IPCE data in the 300-850 nm range at AM1.5G condition ($J_{IPCE}$(MoS$_2$ QDs:f-RGO) > $J_{IPCE}$(f-RGO) > $J_{IPCE}$(Ref.) > $J_{IPCE}$(MoS$_2$ QDs) are in agreement with the $J_{sc}$ of the different PSCs extrapolated by the corresponding I-V curves. The improvement of $V_{oc}$ and FF compared to the PSCs without spiro-OMeTAD is attributed to both the high WF values of the spiro-OMeTAD compared to those of ABLs, which assist the hole-extraction, and the suppression of the contact between MAPbI$_3$ and Au, where charge recombination and/or chemical degradation of MAPbI$_3$ can occur.[244]

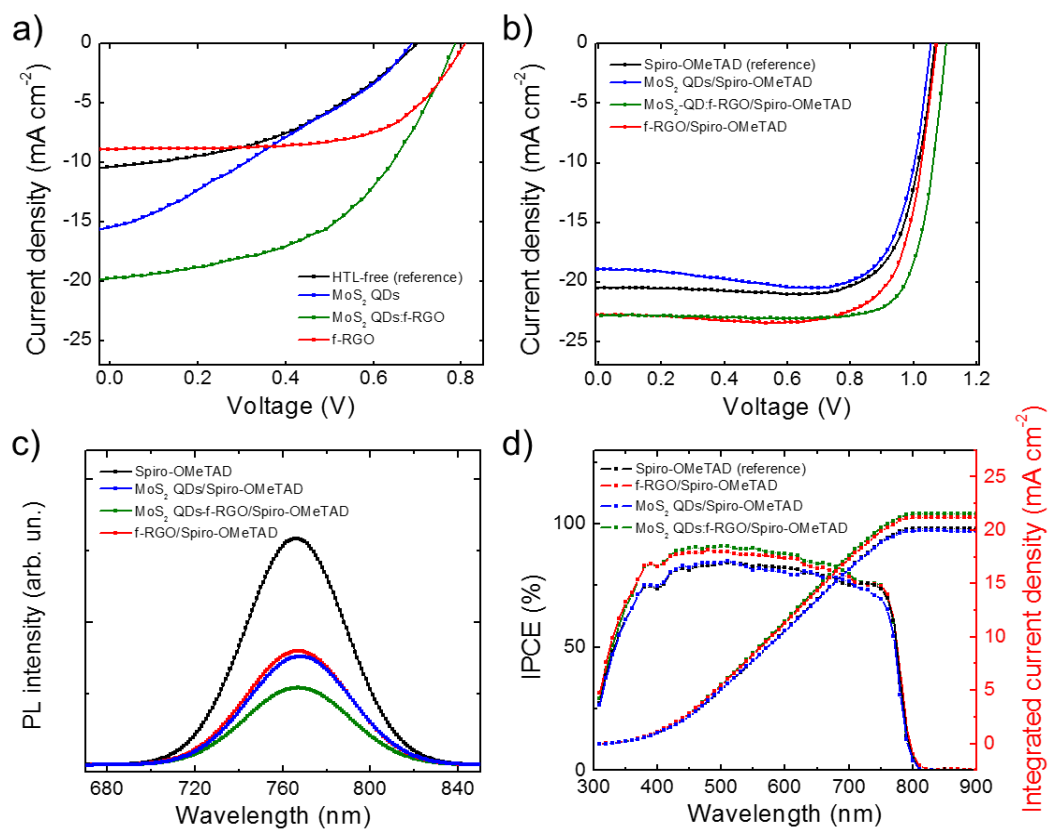

**Figure 5.** (a,b) I-V characteristics of tested PSCs using MoS$_2$ QDs, f-RGO and MoS$_2$ QDs:f-RGO as HTL (panel a) or ABL between MAPbI$_3$ and spiro OMeTAD. The data for HTL-free and ABL-free devices are shown as references. (c) Steady-state PL measurements of the MAPbI$_3$ after deposition of spiro-OMeTAD and ABL/spiro-OMeTAD. (d) Incident power conversion efficiency measurements of the various PSCs. The integrated current density of the curves is also shown on the right y-axis (red color).

The statistical PV Figures of Merit (FoM) measured for each set of PSCs using spiro-OMeTAD as HTL and MoS$_2$ QDs, f-RGO and MoS$_2$ QDs:f-RGO as ABLs are reported in **Figure 6**, in comparison with those obtained for ABL-free PSC (Ref.). **Table 1** summarizes the PV FoM extracted by the I-V curves of the PSCs shown in Figure 6. These results demonstrate the reproducibility of the PV performance of the PSC incorporating the ABLs. In particular, PSCs using MoS$_2$ QDs:f-RGO exhibited average PCE value of 18.86 ± 0.72%, corresponding to an increase of 10.6% compared to the reference device without ABL (average PCE = 17.08 ± 0.73%).

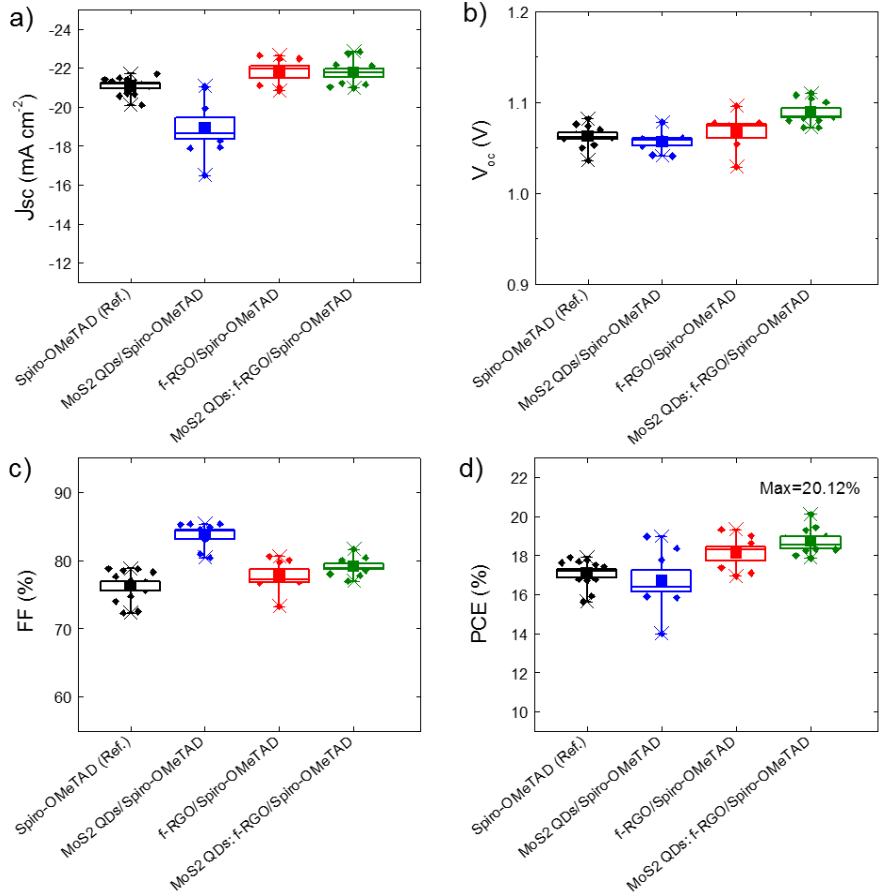

**Figure 6.** Photovoltaic parameters measured at 1 SUN with the relative standard deviation on 12 PSCs for the four investigated PSCs using spiro-OMeTAD as HTL and $MoS_2$ QDs, f-RGO and $MoS_2$ QDs:f-RGO as ABLs: (a) $J_{sc}$; (b) $V_{oc}$; (c) FF and (d) PCE. The average values are indicated by square (■).

**Table 1.** Photovoltaic FOM extracted by the I-V curves of the best performing device of each type of PSCs using spiro-OMeTAD as HTL and $MoS_2$ QDs, f-RGO and $MoS_2$ QDs:f-RGO as ABLs. The average PCE of each type of PCS is also shown.

| Structure of PSC | $V_{oc}$ [V] | $J_{sc}$ [mA cm$^{-2}$] | FF [-] | PCE [%] | Average PCE[a] [%] |
|---|---|---|---|---|---|
| spiro-OMeTAD | 1.06 | 21.49 | 78.31 | 17.53 | 17.08 ± 0.73 |
| $MoS_2$ QDs/spiro-OMeTAD | 1.06 | 20.98 | 83.02 | 18.98 | 16.71 ± 1.60 |
| f-RGO/spiro-OMeTAD | 1.07 | 22.49 | 80.61 | 19.34 | 18.11 ± 0.96 |
| $MoS_2$ QDs:f-RGO/spiro-OMeTAD | 1.11 | 22.81 | 79.75 | 20.12 | 18.76 ± 0.72 |

[a] Average PCE on twelve devices for each type of PSC together with the standard error

In addition to PCE, the long-term stability of PSCs is crucial for real applications.[245] Although $MAPbI_3$ revolutionized the worldwide PV research in the last year, it is intrinsic instable due to its hydroscopicity and tendency to back-convert into its precursors, namely $PbI_2$ and MAI, during

moisture,[246-250] oxygen,[246-248,251-253] and light illumination[246-248,251,254] exposure. In addition, MAPbI$_3$ undergoes a phase transition from the tetragonal to cubic phase at ~54 °C,[255-257] a temperature that can be reached during typical solar cell operation, being not compatible with certification requirement of solar modules (-40–85 °C temperature range).[257] This represents the major constraint for the market breakthrough of this technology.[190,258] So far, the chemical engineering of the perovskite absorber elemental composition has been proved to address the instability issues.[25,259,260] Device lifetime close to market requirements, *e.g.* 500-hours stability and > 20% PCE, has been recently achieved by formulating perovskites with mixed cations *i.e.* formamidinium (FA), MA and inorganic species (Cs or Ru).[261,262] One-year stable PSCs were achieved by engineering an ultra-stable 2D/3D (HOOC(CH$_2$)$_4$NH$_3$)$_2$PbI$_4$/CH$_3$NH$_3$PbI$_3$ perovskite junction.[39] Despite this result, interface engineering of PSCs also affects their stability, [16,60,61,69,263] since the diffusion of elemental species such as iodine (I) and metal from the electrode materials (*e.g.* Au[244] or Ag[264]) has been recently correlated with the degradation of interfaces and the decay of the PV properties.[265] In this context, the incorporation of graphene flakes into mTiO$_2$ has been demonstrated to increase the chemical stability of overlying MAPbI$_3$, which exhibited higher crystalline quality compared to the one of MAPbI$_3$ deposited directly onto mTiO$_2$[69] and a freezed tetragonal phase regardless of the temperature.[69] Active buffer layer based on GRMs improved the charge extraction process compared to that of ABL-free reference, preventing the degradation induced by the diffusion of Au and I. [71,104,231,266,267] Although it was not the goal of our work to overcome the intrinsic instability of MAPbI$_3$, the stability of the device after encapsulation was measured in ISOS-D-1 shelf-life aging test protocol[268] (**Figure 7**). After 1032 h-aging test, the ABL-based PSCs exhibited a decrease of PCE of only 13.5%, 11.2% and 8.8% for MoS$_2$ QDs-, f-RGO- and MoS$_2$ QDs:f-RGO-based PSCs, respectively. These reduction values are significantly lower than the ones shown by the reference PSC without ABL (24.6%).

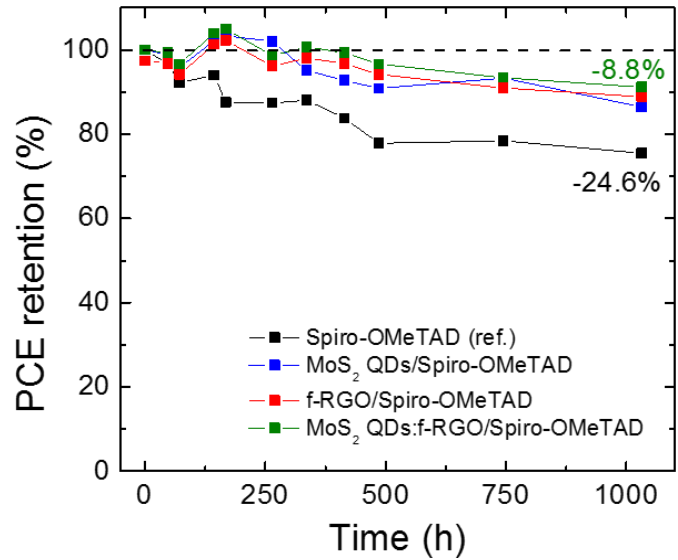

**Figure 7.** Normalized PCE trends *vs.* time extracted by I–V characteristics under 1 SUN illumination, periodically acquired during the shelf life test (ISOS-D-1) for the four PSCs.

The improved stability of the ABL-based PSCs compared to the reference PSC is ascribed to the surface passivation of the perovskite layer provided by the ABLs, which mitigate the I migration from the $MAPbI_3$ into the spiro-OMeTAD[269] and the formation of Au pathways from the metal electrode to the $MAPbI_3$.[270]

## CONCLUSIONS

In conclusion, solution processed low dimensional materials can be designed and combined to improve both efficiency and stability performances of PSCs *via* interface engineering. These results, coupled with the availability of a wide library of 2D materials, demonstrate that GIE is a powerful tool for boosting the PV performance of PSCs. Moreover, 2D materials can be directly produced from cost-effective and environmentally friendly solution-processed methods[78] from their bulk counterparts,[81-84] allowing the formulation of inks with on-demand (opto)electronic properties.[86-88] Solution processed 2D materials can be deposited on different substrates using established printing/coating techniques,[87] in order to be integrated as functional layers (*e.g.* charge transport layers and ABLs) into the PSC structures.[271] By a deep exploitation of the potential offered by 2D materials,[75,199,] we "*ad-hoc*" designed $MoS_2$ QDs anchored to functional site of RGO flakes to effectively collect the photogenerated holes (as well as blocking electron) from $MAPbI_3$ towards the anode contact in mesoscopic MAPbI$_3$-based PSCs, reaching a maximum PCE values of 20.12% (average PCE of 18.8%). The use of 2D materials is also beneficial for the stability increase of mesoscopic PSCs, indicating the feasibility towards next-generation of PSCs, which exploit both GIE and efficient and stable perovskite chemistries, including mixed cation/halide and 2D perovskites.

**EXPERIMENTAL METHODS**

**Production of materials**

Graphene oxide was synthesized from graphite flakes (Sigma Aldrich, +100 mesh ≥75% min) using a modified Hummer's method.[182] Briefly, 1 g of graphite and 0.5 g of NaNO$_3$ (Sigma Aldrich, reagent grade) were mixed, followed by the dropwise addition of 25 mL of H$_2$SO$_4$ (Sigma Aldrich). After 4 h, 3 g of KMnO$_4$ (Alpha Aesar, ACS 99%) was added slowly to the above solution, keeping the temperature at 4 °C with the aid of an ice bath. The mixture was let to react at room temperature overnight and the resulting solution is diluted by adding 2 L of distilled water under vigorous stirring. Finally, the sample was filtered and rinsed with H$_2$O. Finally the sample was dried at 110 °C overnight.

Reduced graphene oxide was produced by thermal reduction of the as-produced GO[180,181] in a quartz tube (120 cm length and 25 mm inner diameter) passing through a three zones split furnace (PSC 12/--/600H, Lenton, UK). Experimentally, gas flows were controlled upstream by an array of mass flow controllers (1479A, mks, USA). Under a 100 sccm flow of Ar/H$_2$ (90/10 %), 100 mg of GO were heated to 100 °C for 20 min to remove the presence of water residuals. Subsequently, a ramp of 20 °C min$^{-1}$ was used to reach 1000 °C, and stabilized at this temperature for 2 h. Finally, the oven was left to cool to room temperature.

The RGO was functionalized with MPTS, (95%, Sigma Aldrich) in an ethanol (absolute alcohol, ≥99.8%, without additive, Sigma Aldrich) solution by reflux at 60°C for 15h.[166] For this reaction 250 μL of MPTS were added per mg of RGO. After the synthesis, the silane-functionalized RGO material was recovered by centrifugation (9000 rpm) and re-dispersed in ethanol (EtOH) by vortexing for a second centrifugation (9000 rpm) to remove unreacted silane. A solvent-exchange process[184,185,95] was carried out to re-disperse the f-RGO in IPA at a concentration of 0.4 mg mL$^{-1}$.

Molybdenum disulfide quantum dots were produced through a one-step solvothermal method starting from MoS$_2$ flakes, produced by LPE[144] of bulk MoS$_2$ crystals in IPA followed by SBS.[172,173] In detail, 30 mg of MoS$_2$ bulk crystal (Sigma Aldrich) were added to 50 mL of IPA and then ultrasonicated (Branson® 5800 cleaner, Branson Ultrasonics) for 8 h. The resulting dispersion was ultracentrifuged (Optima™ XE-90 ultracentrifuge, Beckman Coulter) for 15 min at 2700 $g$, in order to separate the un-exfoliated MoS$_2$ crystals (collected as sediment) from the thinner MoS$_2$ flakes that remain in the supernatant. Then, the sample was refluxed in air under stirring for 24 h at 140 °C. The resulting dispersion was subsequently ultracentrifuged for 30 min at 24600 $g$. Afterward, the supernatant was collected, obtaining the MoS$_2$ QDs dispersion. By evaporating the solvent, a concentration of 0.2 mg mL$^{-1}$ was obtained.

The hybrid dispersion between MoS$_2$ QDs and f-RGO were produced by mixing the as-produced component dispersions in a volume ratio of 1:1 (corresponding to a weight ratio of 1:2). By evaporating the solvent, the concentration was doubled in order to have the same amount of the material compared to the native dispersions.

**Characterization of materials**

Transmission electron microscopy images were taken with a JEM 1011 (JEOL) TEM (thermionic W filament), operating at 100 kV. Morphological and statistical analysis was carried out by using ImageJ® software (NIH) and OriginPro® 9.1 software (OriginLab), respectively. The statistical analysis was performed on 50 flakes from the different TEM images collected. The lateral dimension of each flake was calculated as the maximum Feret's Diameter. Samples for the TEM measurements were prepared by drop casting the material dispersions onto ultrathin carbon-coated copper grids rinsed with deionized water and subsequently dried under vacuum overnight.

Atomic force microscopy images were taken using a Nanowizard III (JPK Instruments, Germany) mounted onto an Axio Observer D1 (Carl Zeiss, Germany) inverted optical microscope. The AFM measurements were carried out by using PPP-NCHR cantilevers (Nanosensors, USA) with a nominal tip diameter of 10 nm. A drive frequency of ~295 kHz is used. Intermittent contact mode AFM images (512×512 data points) of 2.5×2.5 $\mu m^2$ were collected by keeping the working set point above 70% of the free oscillation amplitude. The scan rate for acquisition of images was 0.7 Hz. Height profiles were processed by using the JPK Data Processing software (JPK Instruments, Germany) and the data were analysed with OriginPro® 9.1 software. Statistical analysis was carried out by means of Origin 9.1 software on multiple AFM images for each sample, and calculated on 50 flakes. The samples were prepared by drop-casting the materials dispersions onto mica sheets (G250-1, Agar Scientific Ltd., Essex, U.K.) and dried under vacuum.

Optical absorbtion spectroscopy measurements were carried out on material dispersions by using a Cary Varian 5000 UV–vis spectrometer.

X-ray photoelectron spectroscopy characterization was carried out on a Kratos Axis UltraDLD spectrometer, using a monochromatic Al Kα source (15 kV, 20 mA). The spectra were taken on a 300×700 $\mu m^2$ area. Wide scans were collected with constant pass energy of 160 eV and energy step of 1 eV. High-resolution spectra were acquired at constant pass energy of 10 eV and energy step of 0.1 eV. The binding energy scale was referenced to the C 1s peak at 284.8 eV. The spectra were analysed using the CasaXPS software (version 2.3.17). The samples were prepared by drop-casting the material dispersions onto Si/$SiO_2$ substrate (LDB Technologies Ltd) and dried under vacuum.

Fourier-transform infrared spectroscopy was performed in a Bruker Vertex® 70v (4000-400 cm$^{-1}$ range, 100 scans). The samples were prepared by drop casting MPTS, RGO, f-RGO, and MoS$_2$ QDs:f-RGO films on BaF$_2$ substrates (IR grade, Crystran®, IR open window 4000 to 600 cm$^{-1}$).

Ultraviolet photoelectron spectroscopy analysis was performed to estimate the energy Fermi level ($E_F$) of the materials under investigation with the same equipment used by XPS, and adopting a He I (21.22 eV) discharge lamp. The $E_F$ was measured from the threshold energy for the emission of secondary electrons during He I excitation. A −9.0 V bias was applied to the sample in order to precisely determine the low kinetic energy cutoff. The samples were prepared by drop-casting onto 50 nm Au-sputter-coated silicon wafers.

**Fabrication of solar cells**

The solar cells containing four devices were fabricated on laser patterned glass/FTO substrates (Pilkington, 8 Ω □$^{-1}$), which were washed for 15 min with acetone, ethanol and deionized water in an ultrasonic bath, respectively. Furthermore, a compact 40 nm TiO$_2$ layer (c-TiO$_2$) was deposited onto the pre-cleaned laser patterned FTO glass *via* spray pyrolysis (450°C) from a solution consisting of 0.16 M diisopropoxytitanium bis(acetylacetonate) (Ti(AcAc)$_2$) and 0.4 M acetyl acetone (AcAc) in ethanol. For the mesoporous TiO$_2$ (m-TiO$_2$) layer, anatase TiO$_2$ nanoparticles paste (30NRD, GreatCell SolarDyesol®) were dissolved in ethanol by stirring at a w/w ratio of 1:6. Mesoporous layer was deposited onto c-TiO$_2$ by spin-coating 140μL of paste at 3000 rpm for 15 s, and subsequently sintered at 480 °C for 30 min.

Successively, MAPbI$_3$ perovskite absorber layer was deposited by solvent engineering method. Briefly 717.76 mg mL$^{-1}$ of PbI$_2$ and 247.56 mg mL$^{-1}$ of CH$_3$NH$_3$I were dissolved in dimethylformamide:dimethylsulfoxide (DMF:DMSO) 8:1 (v/v) by stirring for 24h at room temperature to obtain the perovskite-based solution. 70 μL of the perovskite solution was spin coated on the mesoporous layer with two steps spinning, first 1000 rpm for 10 s and then 5000 rpm for 45 s.

Just 34 s before the end of the second spin coating step, 0.7 mL of diethyl ether was dropped on the substrates. Subsequently, the perovskite layer was treated with a double-step annealing process, performed at 50°C for 2 min and then at 100°C for 10 min. After the heat treatment of the perovskite layer, the 2D materials dispersed in IPA were deposited by an automated spray coating equipment (Aurel®) onto perovskite layer by using $N_2$ flow (see SI, for the spray parameter settings, **Figure S12**). 100 μL of HTL material solution containing spiro-OMeTAD (73.5 mg.mL$^{-1}$, Borun® sublimed grade >99.8%) in chlorobenzene (CB, Sigma Aldrich) doped with 26μL of tert-butylpyridine (TBP, Sigma Aldrich, 96%), 16.6 μL of lithium bis(trifluoromethanesulfonyl)imide (Li-TFSI, Sigma Aldrich, 99.95%) of stock solution (520 mg in 1 mL acetonitrile (Sigma Aldrich), and 7.2 μL of cobalt (III) complex solution (FK209 from Lumtec®) was deposited by spin coating at 2000 rpm for 20 s. Finally, 80 nm of Au counter electrode was deposited by thermal evaporation in high vacuum condition ($10^{-6}$ mbar). For the shelf life tests, the device were encapsulated following the protocol previously reported in ref. 269.

**Characterization of solar cells**

Scanning electron microscopy analysis of solar cells was performed using a Helios Nanolab® 600 DualBeam microscope (FEI Company) and 10kV and 0.2 nA as measurement conditions. For the EDX spectra acquisition and analysis on the solar cells we used the microscope combined with an X-Max detector and INCA® system (Oxford Instruments) and 15kV and 0.8 nA as measurement conditions. The samples were imaged without any metal coating or pre-treatment. To evaluate the layered stack of the solar cell by cross section, the samples were prepared using focused ion beam coupled to the microscope.

Current-Voltage (I-V) characteristics of masked and encapsulated devices were acquired in air atmosphere by using a solar simulator (ABET Sun 2000, class A) at AM1.5 and 100 mW cm$^{-2}$ illumination conditions, calibrated with a certified reference Si Cell (RERA Solutions RR-1002). Devices were not preconditioned before the I-V measurements. I-V scans were performed by using a scan rate of 20mV s$^{-1}$. Incident photon to current conversion efficiency spectra acquisition were carried out by means of a home-made setup composed by a monochromator (Newport, mod. 74000) coupled with a Xe lamp (Oriel Apex, Newport) and a source meter (Keithley, mod. 2612). A home-made LabVIEW program controlled the spectra acquisition.

Shelf-life test was carried out on encapsulated devices (by following the indications of the ISOS-D-1 shelf life ageing test protocol.[272] In particular, the devices were kept in the dark, dry conditions (relative humidity < 50%) and at open circuit.

## AUTHOR INFORMATION

### Corresponding Author

* Tel: +39 01071781795. E-mail: francesco.bonaccorso@iit.it

* Tel: +39 0672597456. E-mail: aldo.dicarlo@uniroma2.it.

### Author Contributions

The manuscript was written through contributions of all authors. All authors have given approval to the final version of the manuscript. ‡These authors contributed equally.

### Funding Sources

This project has received funding from the European Union's Horizon 2020 research and innovation program under grant agreement no.785219-GrapheneCore2.

## ACKNOWLEDGMENT

We thank Electron Microscopy facility – Istituto Italiano di Tecnologia for support in TEM data acquisition; and IIT Clean Room facility and Smart Materials Group for the access to carry out SEM/EDS measurements and FTIR characterization, respectively. ADC gratefully acknowledges the financial support of the Ministry of Education and Science of the Russian Federation in the framework of Megagrant N° 14.Y26.31.0027.

# Supporting Information

**Morphological analysis of MoS₂ flakes.**

**Figure S1** reports the morphological analysis (*i.e.*, lateral size and thickness) of the liquid phase exfoliated MoS$_2$ flakes, from which MoS$_2$ quantum dots (QDs) were derived by subsequent solvothermal treatment. Figure S1a shows a representative transmission electron microscopy (TEM) image of the MoS$_2$ flakes, which exhibited regular shaped borders. Figure S1b reports the statistical analysis of lateral dimension of the MoS$_2$ flakes, which exhibited an average lateral size of ~420 nm. Figure S1c shows a representative atomic force microscopy (AFM) image of the MoS$_2$ flakes. Height profiles (dashed white lines) indicated the presence of one- and two-layer flakes (the monolayer thickness is between 0.7 and 0.8 nm[1,2]). Figure S1d shows the statistical analysis of the thickness of MoS$_2$ flakes, which exhibited an average thickness of ~2.7 nm.

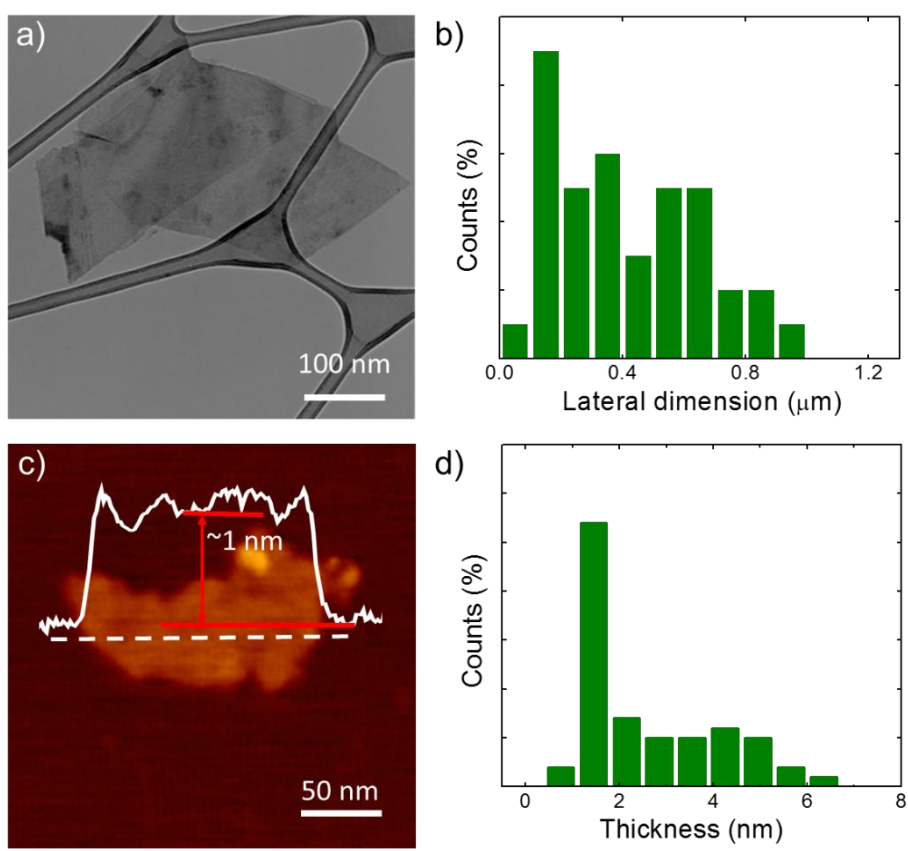

**Figure S1.** Morphological characterization of the as-produced MoS$_2$ flakes. (a) TEM image of the MoS$_2$ flakes. (b) Statistical analysis of the lateral dimension of the MoS$_2$ flakes (calculated on 100

flakes from different TEM images). (c) AFM image of the MoS$_2$ flakes. Representative height profile (solid white lines) of the indicated section (white dashed lines) is also shown. (d) Statistical analysis of the thickness of the MoS$_2$ flakes (calculated on 50 flakes from different AFM images).

**Morphological analysis of RGO flakes**

**Figure S2** reports the morphological analysis of the as-produced reduced graphene oxide (RGO) flakes. Figure S2a shows a representative TEM image of the RGO flakes, which have an irregular shape and rippled structure. Figure S2b reports the statistical analysis of the lateral dimension of the RGO flakes, which have an average value of 1.7 µm. Figure S2c shows a representative AFM image of the RGO flakes. Height profiles (dashed white lines) evidence nano-edge steps between 0.6 and 1.6 nm. Figure S2d shows the statistical analysis of the RGO flakes, which have an average thickness of ~1.8 nm. This indicates that few-layer RGO flakes were effectively produced (the monolayer thickness is ~0.34 nm[3,4]).

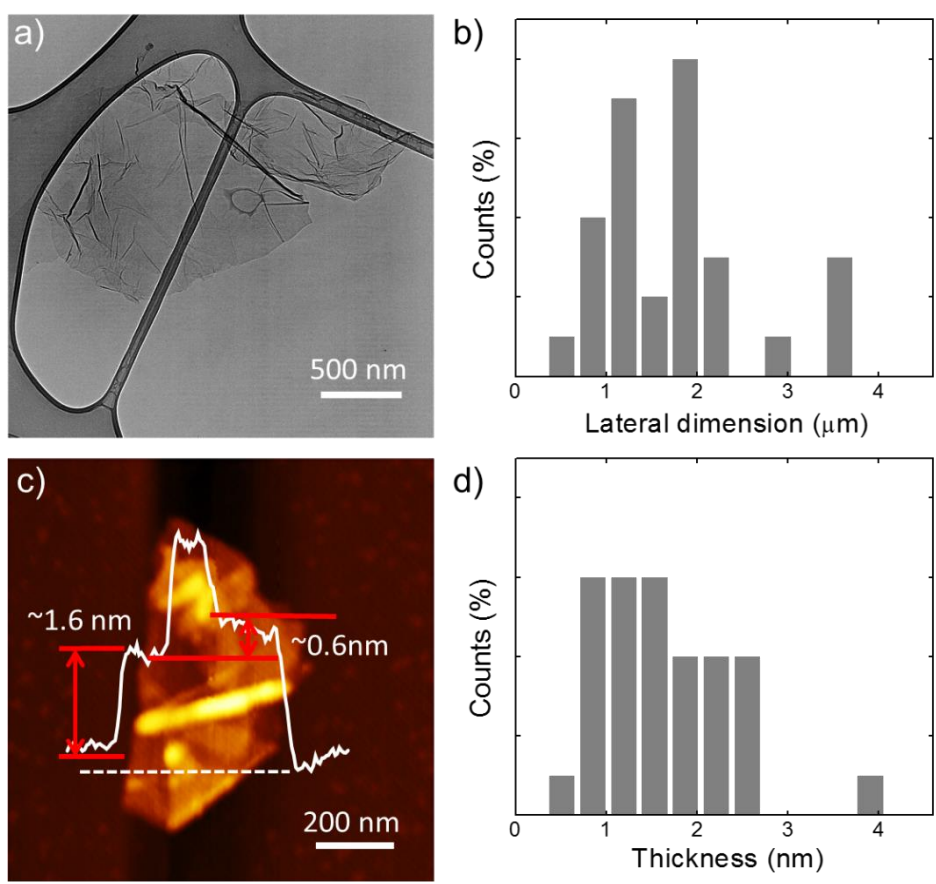

**Figure S2.** Morphological characterization of the as-produced RGO flakes. (a) TEM image of the RGO flakes. (b) Statistical analysis of the lateral dimension of the RGO flakes (calculated on 100

flakes from different TEM images). (c) AFM image of the RGO flakes. Representative height profile (solid white lines) of the indicated section (white dashed lines) is also shown. (d) Statistical analysis of the thickness of the RGO flakes (calculated on 50 flakes from different AFM images).

**Raman spectroscopy characterization of materials**

**Figure S3**a reports representative Raman spectra of $MoS_2$ QDs, compared to both native bulk $MoS_2$ powder and $MoS_2$ flakes. These spectra show the presence of first-order modes at the Brillouin zone center $E_{2g}^1(\Gamma)$ (~379 cm$^{-1}$ for both $MoS_2$ flakes and QDs, and ~377 cm$^{-1}$ for bulk $MoS_2$) and $A_{1g}(\Gamma)$ (~403 cm$^{-1}$), involving the in-plane displacement of Mo and S atoms and the out-of-plane displacement of S atoms, respectively.[5,6] The $E_{2g}^1(\Gamma)$ mode of both the $MoS_2$ flakes and QDs exhibits softening compared to the one of the bulk $MoS_2$. The shift of the $E_{2g}^1(\Gamma)$ mode is explained by the dielectric screening of long-range Coulomb $MoS_2$ interlayer interaction.[5] The full width at half maximum (FWHM) of the $E_{2g}^1(\Gamma)$ and $A_{1g}(\Gamma)$ (*i.e.*, FWHM($E_{2g}^1(\Gamma)$) and FWHM($A_{1g}(\Gamma)$), respectively) of both $MoS_2$ flakes and $MoS_2$ QDs increases of ~3 cm$^{-1}$ and ~2 cm$^{-1}$, respectively, compared to the corresponding modes of bulk $MoS_2$. The increase of FWHM($A_{1g}(\Gamma)$) for $MoS_2$ flakes and $MoS_2$ QDs is attributed to the variation of interlayer force constants between the inner and outer layers.[2]

Figure S3b shows representative Raman spectra of functionalized RGO (f-RGO) flakes, together with its native materials, *i.e.*, graphene oxide (GO) and RGO flakes. The Raman spectrum of GO flakes reveals two main peaks located at 1352 and 1591 cm$^{-1}$, corresponding to D and G bands, respectively.[7,8] The G peak corresponds to the $E_{2g}$ phonon at the Brillouin zone center,[7,8] while the D peak is due to the breathing modes of sp$^2$ rings,[7,8] requiring a defect for its activation by double resonance.[7] The 2D peak position, located at ~2700 cm$^{-1}$ is the second order of the D peak.[9] Double resonance can also happen as an intravalley process, *i.e.*, connecting two points belonging to the same cone around K or K'.[9] This process gives rise to the D' peak, which is usually located at ~1600 cm$^{-1}$ in presence of high density defects.[9] In these conditions, the D' band is merged with the G band. The 2D' peak, located at ~3200 cm$^{-1}$, is the second order of the D',[9] while D+D', positioned at ~2940 cm$^{-}$

[1] is the combination mode of D and D'. These three peaks show a low intensity, due to electronic scattering,[10] and a very broad line shape. The full width half maximum (FWHM) of D (FWHM(D)) is 127 cm$^{-1}$, while FWHM(G) is 79 cm$^{-1}$. The FWHM(G) always increase with disorder and, indeed, it is much larger than pristine graphene (FWHM(G) < 20cm$^{-1}$)[8] and edge-defected graphene flakes (FWHM(G) ~ 25cm$^{-1}$).[10] The high intensity ratio between the intensity of D and G ($I_D/I_G$) (~0.86) and the large FWHM(D) (~125cm$^{-1}$) is due to the presence of both structural defects (due to oxidation process) and covalent bonds (*e.g.*, C–H, C–O), both contributing to the D peak. In the case of the RGO flakes, the D and G peaks are located at 1352 cm$^{-1}$ and 1597 cm$^{-1}$, respectively, while FWHM(G) and FWHM(D) are 64 cm$^{-1}$ and 83 cm$^{-1}$, respectively. The softening of the G band compared to that of GO flakes could be ascribed to the presence of defected regions as consequence of thermal stresses upon annealing.[11] FWHM(D) and FWHM(G) are narrower compared to those of GO flakes, indicating a restoration of the sp$^2$ rings.[7] The $I_D/I_G$ for RGO (~1.25) is considerably higher than that of GO flakes (~0.86). In fact, $I_G$ is constant as a function of disorder because is related to the relative motion of sp$^2$ carbons,[7] while an increase of $I_D$ is directly linked to the presence of sp$^2$ rings.[7,8] Thus, an increase of the $I_D/I_G$ means the restoration of sp$^2$ rings.[7] For f-RGO flakes, FWHM(D) and FWHM(G) are further reduced compared to those of the RGO flakes, which means that the sp$^2$ rings are preserved.[7] The $I_D/I_G$ decreases compared to both those of GO and RGO flakes. This effect can be ascribed to the edge/defect passivation of RGO flakes after the functionalization process.[7,8]

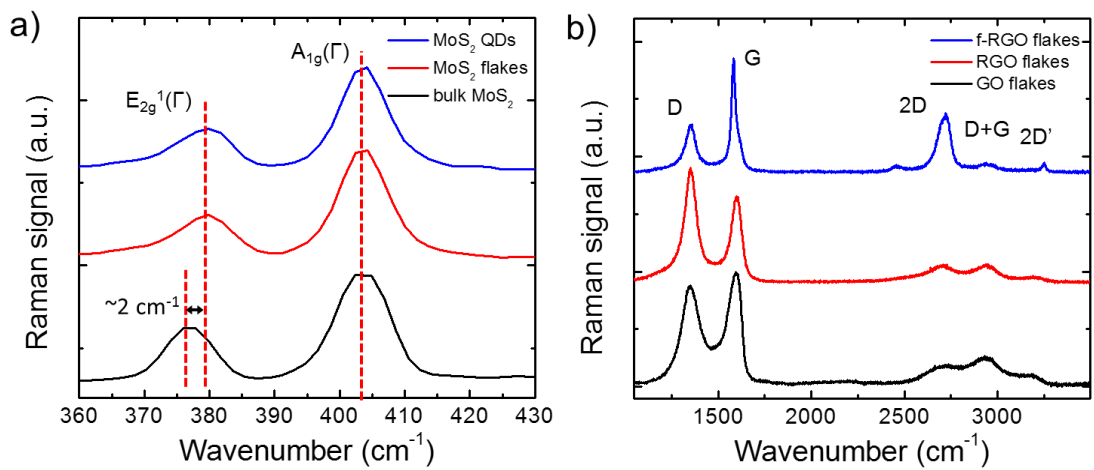

**Figure S3.** Raman spectra: (a) bulk $MoS_2$, $MoS_2$ flakes and $MoS_2$ QDs; (b) GO flakes, RGO flakes and f-RGO flakes.

**Fourier-transform infrared spectroscopy analysis**

Fourier-transform infrared (FTIR) spectroscopy was performed on (3-mercaptopropyl)trimethoxysilane (MPTS), RGO, f-RGO, and MoS$_2$ QDs:f-RGO (**Figure S4**). From the FTIR spectrum of MPTS (**Figure S4**a), it is possible to identify the bands corresponding to Si-O-Si stretching (1089 cm$^{-1}$), S-H stretching (2564 cm$^{-1}$) and C-H stretching (2830/2944 cm$^{-1}$). In agreement with the Scheme 1 of the main text, the first step of the process was the MPTS functionalization of the RGO flakes. As result in the FTIR spectrum of the RGO appeared a broad Si-O-Si stretching band at 1078 cm$^{-1}$ superimposed to the RGO FTIR spectrum (Figure S4b,c). This observation confirms the hydrolyzation and condensation between the oxygen functionalities of the RGO and the alkoxysilane groups (–OCH$_3$) during the MPTS functionalization.[12–14] Notably, the weak peak related to the S-H bond is only shown in the FTIR spectrum of the pure MPTS. This agrees with other studies on the silane functionalization of oxide nanoparticles.[15] In the FTIR spectrum of the MoS$_2$ QDs:f-RGO, the broad Si-O-Si stretching band of the f-RGO is still at 1078 cm$^{-1}$. This indicates that the hybridization did not alter the MPTS functionalization of the RGO and, at the same time, that the interaction between MoS$_2$ QDs and f-RGO only took place at the free SH groups of the MPTS.

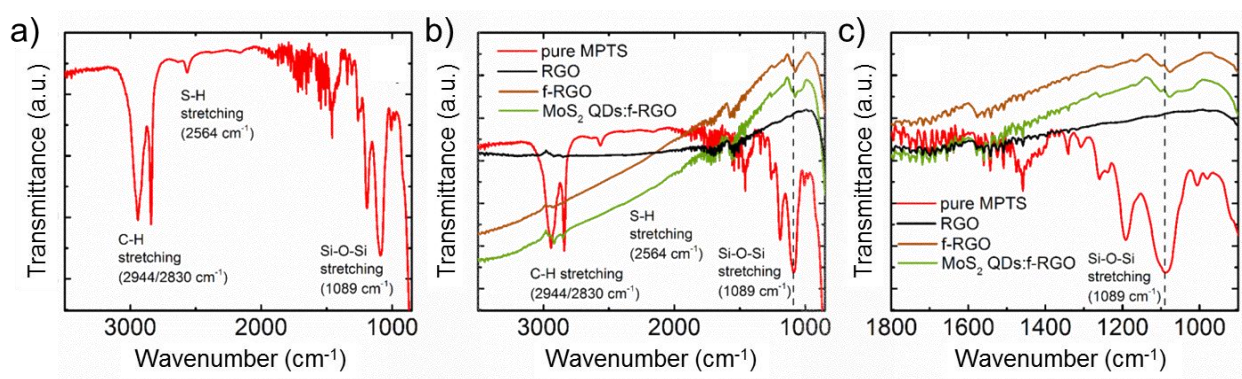

**Figure S4.** (a) FTIR spectrum of the pure MPTS indicating the main characteristic features: Si-O-Si, S-H and C-H stretching vibrations. (b) Comparison of the FTIR spectra corresponding to pure MPTS, RGO, f-RGO and MoS$_2$ QDs:f-RGO materials. (c) Corresponding zoom in the Si-O-Si area for the spectra shown in (b).

**Gravitational sedimentation of the RGO and f-RGO dispersion in ethanol**

**Figure S5** shows photographs of 1 mg mL$^{-1}$ RGO and f-RGO dispersions in ethanol (EtOH) after 2 h of gravitational sedimentation. The photographs show a clear sedimentation of the dispersion of RGO due to the poor hydrogen-bonding capability of the as-produced material. After the functionalization of the RGO, the presence of MPTS groups decreased the surface energy of the flakes, making f-RGO compatible with polar solvent, such as EtOH. Consequently, no significant gravitational sedimentation for f-RGO dispersions in EtOH was observed in the same timeframe.

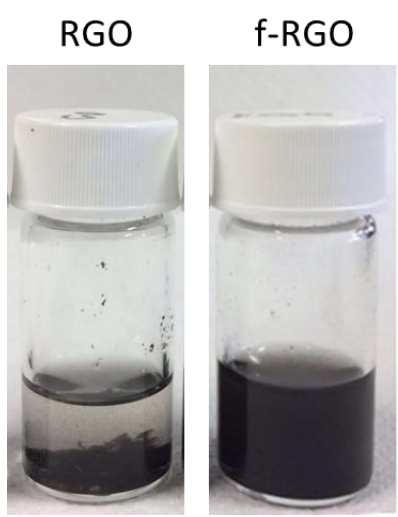

**Figure S5.** Photograph of 1 mg mL$^{-1}$ RGO and f-RGO dispersions in EtOH after 2 h of gravitational sedimentation.

**Photoluminescence analysis of MoS$_2$ QDs**

**Figure S6** reports the photoluminescence (PL) spectra of MoS$_2$ QDs dispersion in 2-Propanol (IPA), collected at different excitation wavelengths (from 300 to 500 nm). The PL peaks were red-shifted with the increase of the excitation wavelength. This excitation-dependent PL emission can be ascribed to quantum confinement and edge state emission effect.[18-20] The sharp small features observed in the spectra were related to the IPA solvent, as proved by blank PL spectrum (inset to Figure S6).

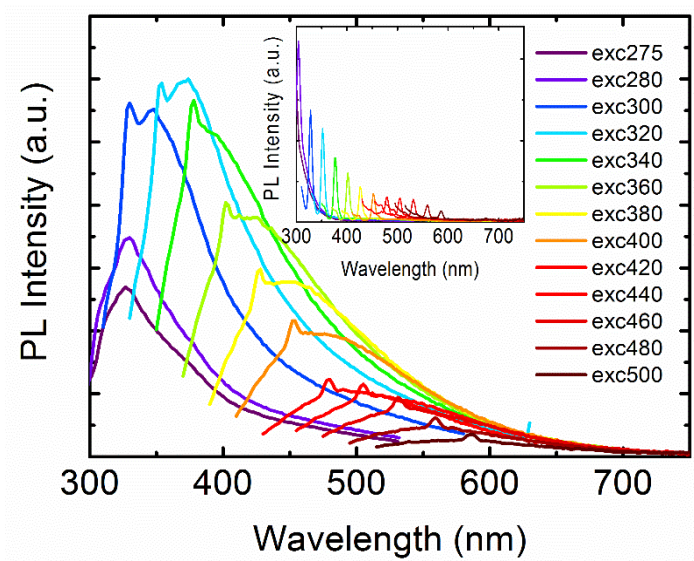

**Figure S6.** Photoluminescence spectra of the MoS$_2$ QDs at different excitation wavelength, ranging from 300 to 500 nm. The inset shows the blank (control) PL spectra of IPA at different excitation wavelengths.

**Supplementary Tauc analysis of MoS₂ QDs**

As commented in the main text, the Tauc plot analysis of nanocrystals can be trivial and the $E_g$ values of MoS$_2$ QDs calculated by Tauc analysis has to be considered qualitatively.[21] Previous work suggested to correct the Tauc relation for direct-allowed transition in nanocrystals by assuming the power factor (*n*) equal to 1.[21] **Figure S7** reports the Tauc plot for *n* =1, which the $E_g$ values to be estimated at ~3.2 eV.

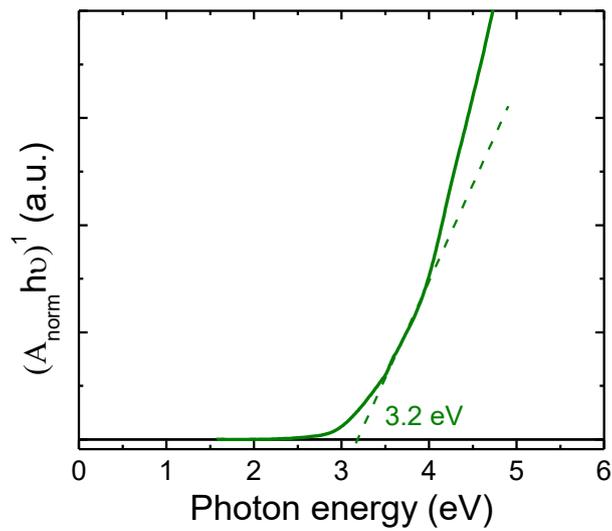

**Figure S7.** Tauc plot of MoS$_2$ QDs with *n* = 1.

**Energy-dispersive X-ray spectroscopy of the MAPbI$_3$/MoS$_2$ QDs:f-RGO**

Scanning electron microscopy combined with energy-dispersive X-ray spectroscopy (SEM/EDX) was carried out in order to gain insight into the coverage of the mesoscopic methylammonium lead iodide (CH$_3$NH$_3$PbI$_3$) perovskite (MAPbI$_3$) layer with MoS$_2$ QDs and f-RGO (**Figure S8**a-c) by preparing the perovskite solar cells (PSCs) as detailed in the Experimental Section (main text) up to the layer object of study. By analyzing the Pb (M, 2.34 keV), Mo (Lα, 2.29 keV) and C (Kα, 0.28 keV) peak signals by INCA® software (XPP matrix correction routine, which considers the atomic number and absorption effects for a standardless estimation of the composition), we confirmed the presence of MoS$_2$ QDs and f-RGO on the PSCs layer stack (Figure S8d-h). For the carefully interpretation of the EDX data, it is important to note that (i) MAPbI$_3$ is already giving a C signal corresponding to the CH$_3$NH$_3$ and (ii) the Pb and Mo signals almost overlap making difficult to discern both components. Moreover, since the spiro-OMeTAD could also contribute to the C signal, we suppressed this layer in the preparation of the devices for these SEM/EDX measurements. As a control, we also determined the I/Pb atomic ratio (I (Lα, 3.94 keV)) in all the samples. In this way, we obtained by evaluating at least 5 areas of about 12×12 μm$^2$ in two samples of each type of PSC values of C/Pb = 2.2 ± 0.1 and I/Pb = 3.0 ± 0.1 (for the reference MAPbI$_3$, Figure S8d); C/Pb = 16 ± 2 and I/Pb = 3.0 ± 0.1 (for f-RGO/MAPbI$_3$, Figure S7e); C/(Pb+Mo) = 2.3 ± 0.2 and I/Pb = 3.0 ± 0.1 (for MoS$_2$ QDs/MAPbI$_3$, Figure S7f) and C/(Pb+Mo) = 15 ± 2 and I/Pb = 3.1 ± 0.1 (for MoS$_2$ QDs:f-RGO/MAPbI$_3$, Figure S8g-h). These results show that the presence of f-RGO promotes the increase of the C/Pb ratio, while the I/Pb stoichiometry (3:1) of the perovskite is preserved.

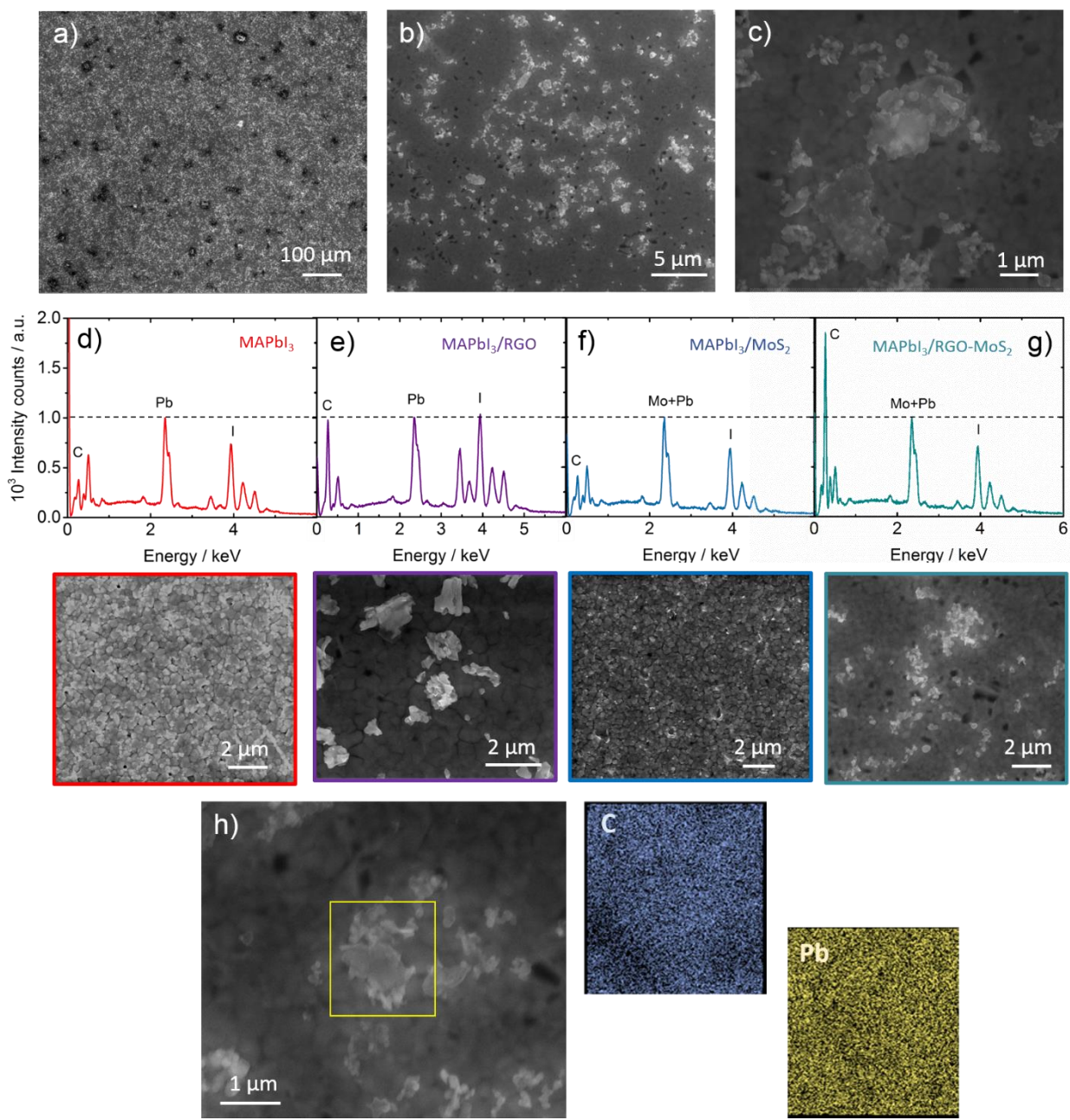

**Figure S8.** (a-c) Representative SEM images at different magnifications for the MoS$_2$ QDs:f-RGO/MAPbI$_3$ sample showing the coverage of the perovskite layer. Representative EDX spectra collected in areas of about 12x12 μm$^2$ for: (d) the reference MAPbI$_3$, (e) f-RGO/MAPbI$_3$, (f) MoS$_2$ QDs/MAPbI$_3$ and (g) MoS$_2$ QDs:f-RGO/MAPbI$_3$ normalized to the Pb signal accompanied of detailed SEM images. (h) A representative SEM/EDX mapping example for a small area (2x2 um$^2$) of MoS$_2$ QDs:f-RGO/MAPbI$_3$ sample pointed out in a yellow square.

**Optical absorption spectroscopy measurements of PSCs**

**Figure S9** shows the UV-VIS absorption spectra of the different PSC architectures before the Au contact deposition. These results evidenced that the optical absorption of the $MoS_2$ QDs:f-RGO-based PSCs did not show significant differences compared to that of $MoS_2$ QDs-based PSC, and exhibited an increase by only ~3% and ~8% compared to that of f-RGO-based and reference PSCs. On the basis of these results, the enhanced $J_{sc}$ value obtained with the use of f-RGO and $MoS_2$ QDs:f-RGO compared to that of reference device and the one of the device based on $MoS_2$ QDs can be attributed to the efficient charge collection in presence of f-RGO and $MoS_2$ QDs:f-RGO, respectively.

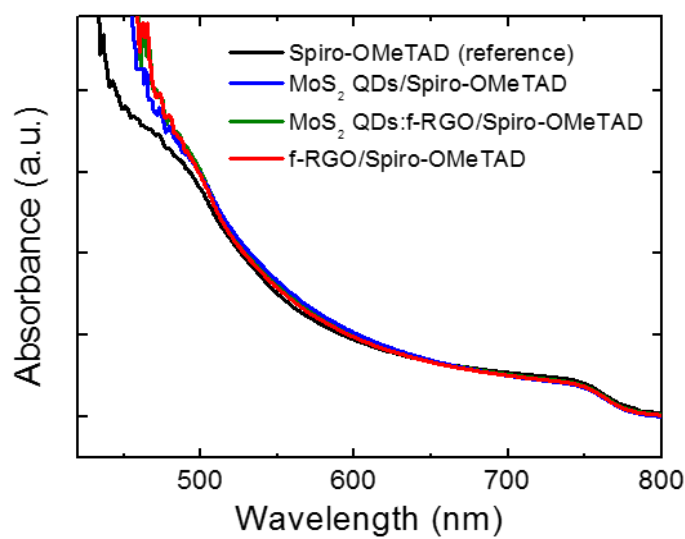

**Figure S9.** Absorption spectra of different PSCs before Au deposition.

**Stabilized power conversion efficiency measurements**

**Figure S10** shows the stabilized power conversion efficiency (PCE) measurement performed with maximum power point (MPP) tracking on PSCs based on $MoS_2$ QDs, f-RGO and $MoS_2$ QDs:f-RGO as active buffer layer (ABL). The comparison with the measurement performed on reference ABL-free PSC based on solely spiro-OMeTAD as hole transport layer (HTL) is also shown. The results confirmed that the presence of $MoS_2$ QDs:f-RGO, as well as the f-RGO, enhance the performance of reference PSC. These results are in agreement with the I-V curve measurements of the various PSCs and the corresponding statistical analysis of the main PV parameters reported in the main text (Figure 5b and Figure 6, respectively).

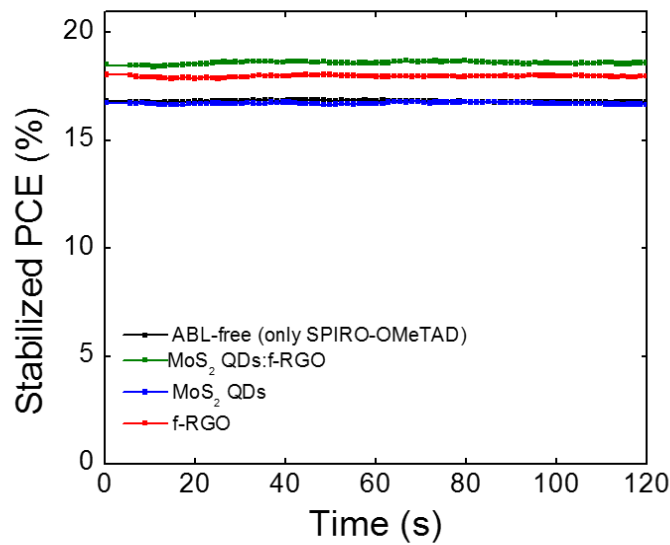

**Figure S10.** Stabilized PCE measurements performed with MPP tracking of the various PSCs based on $MoS_2$ QDs, f-RGO and $MoS_2$ QDs:f-RGO as ABL. The comparison with the measurement performed on reference ABL-free PSC based on exclusively spiro-OMeTAD as HTL is also shown

**Hysteresis analysis**

**Figure S11** reports the forward and reverse I-V curves for the PSCs based MoS$_2$ QDs, f-RGO and MoS$_2$ QDs:f-RGO as ABLs. The comparison with the measurements performed on reference ABL-free PSC based on solely spiro-OMeTAD as HTL is also shown. Clearly, the presence of MoS$_2$ QDs:f-RGO as ABLs, in addition to increase the PCE, also decreases the hysteresis phenomena compared to those of reference PSCs and the other ABL-based devices.

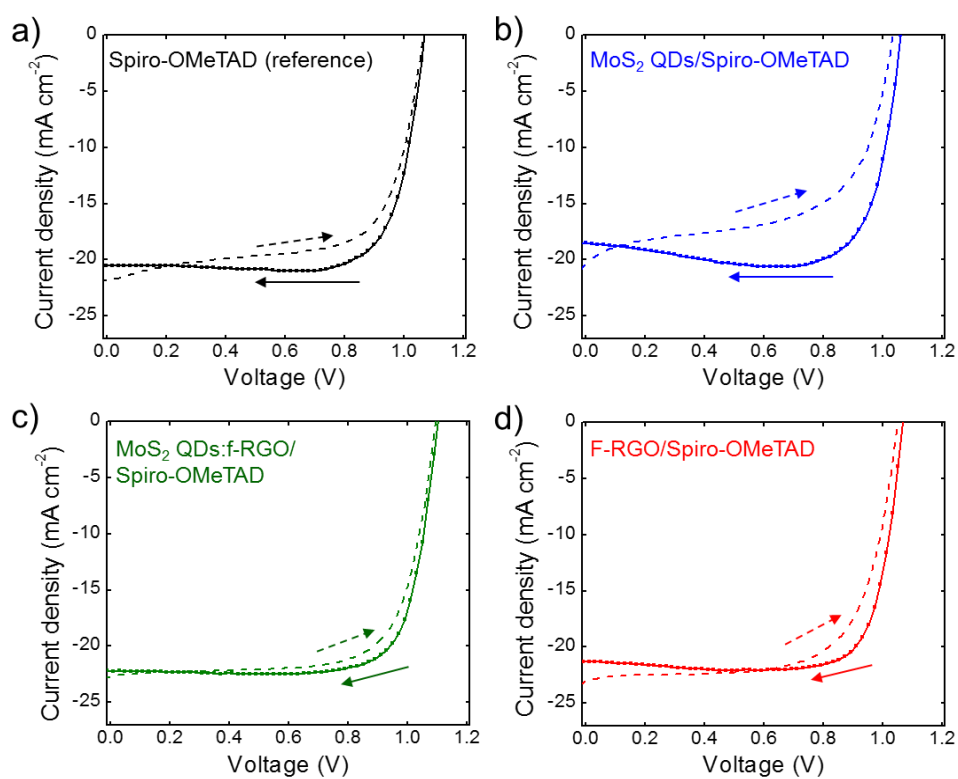

**Figure S11.** Forward and reverse I-V curves for the PSCs based on (a) solely spiro-OMeTAD (reference), (b) spiro-OMeTAD/MoS$_2$ QDs, (c) spiro-OMeTAD/MoS$_2$ QDs:f-RGO and (d) spiro-OMeTAD/f-RGO.

**Spray coating parameter optimization for MoS$_2$ QDs, f-RGO and MoS$_2$ QDs:f-RGO film deposition**

The spray coating deposition of MoS$_2$ QDs, f-RGO and MoS$_2$ QDs:f-RGO dispersion in IPA onto MAPbI$_3$ surface could degrade the native properties of the latter. In order to retain the optical properties of MAPbI$_3$, the spray coating was carried out by using N$_2$ as inert flowing gas, and the deposition parameters (N$_2$ pressure (P$_{N2}$), nozzle aperture (AP$_{nozzle}$), nozzle distance (d$_{nozzle}$), temperature of substrate (T$_{sub}$), spray velocity (v$_{spray}$), and flow rate of solution (FR$_{solution}$)) were optimized by monitoring their influence on the UV-Vis absorption spectrum of MAPbI$_3$ substrates. The degradation of the perovskite was quantified by the relative optical absorption loss, *i.e.,* the module of ratio between the difference of its optical absorption before (Abs$_0$) and after (Abs$_{spray}$) material spray coating referred to Abs$_0$ (in formula: $|(Abs_0-Abs_{spray})/Abs_0)|$).

**Figure S12**a shows the degradation (average values calculated in the ranges of 420-600 nm and 700-780 nm) measured by varying the most critical spray coating parameters and adopting d$_{nozzle}$ = 9 cm, v$_{spray}$ = 300mm s$^{-1}$, FR$_{solution}$ = 20%. The best parameters were found to be: P$_{N2}$ = 1 bar, AP$_{nozzle}$ = 0.6 mm, d$_{nozzle}$ = 9 cm, T$_{sub}$ = 80 °C, v$_{spray}$ = 300mm s$^{-1}$, and a FR$_{solution}$ = 20%. The corresponding degradation of MAPbI$_3$ film was less than 2% (Figure S12b).

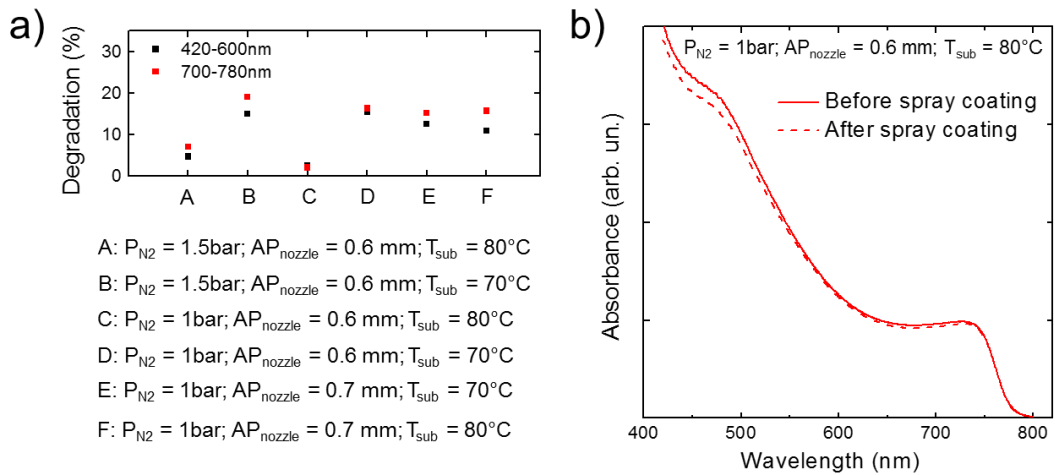

**Figure S12.** a) Sketch of the degradation of MAPbI$_3$ as a function of spray coating settings. b) Absorbance of MAPbI$_3$ before and after spray coating of IPA with optimized parameter setting denoted as C in panel a: $P_{N2}$ = 1 bar, $AP_{nozzle}$ = 0.6 mm, $d_{nozzle}$ = 9 cm, $T_{sub}$ = 80 °C, $v_{spray}$ = 300mm s$^{-1}$, and a $FR_{solution}$ = 20%.